\documentstyle[12pt,epsf]{article}

\def\mathrm#1{{\rm #1}}

\def\Slash#1{#1\hskip-.65em/\,}
\def\slash#1{#1\hskip-.45em/}
\def\mynorm#1{\frac{#1}{16\pi^2}}

\def\wbG/{would-be Goldstone boson}

\renewcommand{\L}{\mathrm{L}}

\newcommand{\R}{\mathrm{R}}
\newcommand{\LR}{\mathrm{LR}}
\newcommand{\gfrac}{\frac{g_3^2}{8\pi^2}}
\newcommand{\loopfac}{\frac{1}{16\pi^2}}
\newcommand{\mb}{m_\mathrm{b}}
\newcommand{\mt}{m_\mathrm{t}}
\newcommand{\mW}{m_\mathrm{W}}
\newcommand{\mH}{m_{\mathrm{H}^\pm}}
\newcommand{\Qb}{Q_\mathrm{b}}
\newcommand{\Qt}{Q_\mathrm{t}}
\newcommand{\QtQb}{\frac{\Qt}{\Qb}}
\newcommand{\xth}{x_\mathrm{tH}}
\newcommand{\xtw}{x_\mathrm{tW}}
\newcommand{\DeltaphiW}{\Delta_{(\phi,W)}}
\newcommand{\DeltaH}{\Delta_{(H)}}
\newcommand{\DeltaHsub}{\Delta'_{(H)}}
\newcommand{\DeltaphiWsub}{\Delta'_{(\phi,W)}}
\newcommand{\Deltachi}{\Delta_{(\chi_j)}}
\newcommand{\Deltaq}{\Delta_{(\tilde u_a)}}
\newcommand{\simless}{\vbox to 1ex{\vss\hbox{
                        $\stackrel{{\scriptstyle <}}{{\scriptstyle\sim}}$
                     }\vss}}

\newcommand{\btosg}{b \to s \gamma}
\newcommand{\cHiggs}{{\mathrm{H}^\pm}}
\newcommand{\GeV}{\mathrm{GeV}}
\newcommand{\eps}{\epsilon}
\newcommand{\jab}{{ja\mathrm{b}}}
\newcommand{\jas}{{ja\mathrm{s}}}
\newcommand{\muh}{\mu_\mathrm{h}}
\newcommand{\myhalf}{\frac{1}{2}}
\newcommand{\third}{\frac{1}{3}}
\newcommand{\quarter}{\frac{1}{4}}
\newcommand{\diag}{\mathrm{diag}}

\newcommand{\Oone}{{O_\LR^1}}
\newcommand{\Otwo}{{O_\LR^2}}
\newcommand{\Othree}{{O_\LR^3}}

\newcommand{\Poneone}{{P^{1,1}_\L}}
\newcommand{\Ponetwo}{{P^{1,2}_\L}}
\newcommand{\Ponethree}{{P^{1,3}_\L}}
\newcommand{\Ponefour}{{P^{1,4}_\L}}

\newcommand{\Ptwo}{{P_\L^2}}
\newcommand{\Pthree}{{P_\L^3}}
\newcommand{\Pfour}{{P_\L^4}}

\newcommand{\Rone}{{R_\L^1}}
\newcommand{\Rtwo}{{R_\L^2}}
\newcommand{\Rthree}{{R_\L^3}}
\newcommand{\Ronel}{{\tilde R_\L^{1,a}}}
\newcommand{\Rtwol}{{\tilde R_\L^{2,a}}}
\newcommand{\Rthreel}{{\tilde R_\L^{3,a}}}
\newcommand{\Rfourl}{{\tilde R_\L^{4,a}}}
\newcommand{\Rfivel}{{\tilde R_\L^{5,a}}}
\newcommand{\Rsixl}{{\tilde R_\L^{6,a}}}

\begin{document}

\begin{titlepage}

\renewcommand{\thefootnote}{\fnsymbol{footnote}}

\rightline{\vbox{\halign{&#\hfil\cr
&SLAC-PUB-6525\cr
&hep-ph/9406286\cr
&June 1994\cr
&T/E\cr
}}}
\vspace{0.8in}

\begin{center}

% \title{

{\Large\bf
On the QCD corrections to $\btosg$ in supersymmetric models%
\footnote{Work supported in part by the Department of Energy under
contract No.~DE--AC03--76SF00515.}
}

\medskip

% \author{

{\large Harald Anlauf%
\footnote{Supported by a fellowship of the Deutsche Forschungsgemeinschaft.}
}
\\ \smallskip
{\it {Stanford Linear Accelerator Center \\
      Stanford University, Stanford, CA 94309}
}\\
\end{center}

\vskip0.30in

% ----------------------------------------------------------------------

\begin{abstract}
We reinvestigate the leading QCD corrections to the radiative decay
$\btosg$ for supersymmetric extensions of the Standard Model.  Although
the major contributions to the corrections originate from the running of
the effective Lagrangian from the W scale down to the $b$ scale,
additional corrections are expected from large mass splittings between
the particles running in the loops, as well as from integrating out
heavy particles at scales different from the W mass.  The calculation is
performed in the framework of effective field theories.
\end{abstract}

\vfill
\begin{center}
  {\em Submitted to Nuclear Physics B}
\end{center}
\vfill

\renewcommand{\thefootnote}{\arabic{footnote}}
\setcounter{footnote}{0}
\end{titlepage}

%%%%%%%%%%%%%%%%%%%%%%%%%%%%%%%%%%%%%%%%%%%%%%%%%%%%%%%%%%%%%%%%%%%%%%%%

\section{Introduction}

Among the rare decays of B mesons, the recently observed radiative weak
decays $\mathrm{B} \to X_s \gamma$ \cite{CLEO93}, where $X_s$ is a
hadronic state with total strangeness $S=-1$, have received much
attention.  As a loop-induced FCNC process it is in particular sensitive
to contributions from physics beyond the Standard Model (SM).

Since the $b$ quark mass is much larger than the QCD scale $\Lambda$,
one assumes that the inclusive decay rate is well described by the
spectator model, where the $b$ quark undergoes a radiative decay.  The
transition amplitude is given by the matrix element an effective
magnetic moment operator.  To lowest order \cite{IL81}, the coefficient
of this operator is obtained by integrating out all heavy particles ($t$
quark, W boson, \ldots), leaving one with an effective field theory
describing the transition $\btosg$ at the parton level at the weak
scale.  The QCD corrections to this coefficient%
\footnote{For a recent review and earlier work see
e.g.~ref.~\cite{GOSN93}.}
have been calculated to leading logarithmic accuracy in
\cite{GSW90,GOSN88,CCRV90,Mis91,AY93,CFMRS93} and are known to enhance
the rate within the SM by a factor of 2--4, depending on the masses of
the $b$ and of the $t$ quark.  This enhancement is, however, subject to
large uncertainties due to the poor knowledge of some input parameters
like the strong coupling constant $\alpha_S$ and due to the residual
renormalization scale dependence (for a recent discussion in the context
of $\btosg$ see e.g.~\cite{AG93,Bor93,BMMP93}), which we will however
not address in this work.

On the other hand, if the particles in the loop have vastly different
masses, one expects sizeable corrections to the Wilson coefficients
already at the weak scale.  These contributions, which are usually
considered as a next-to-leading order effect, have been discussed in
ref.~\cite{CG91} for the Standard Model where, in the case of a top
quark much heavier than the W, they were found to give an additional
enhancement of the order of 20\%.

In the case of the Minimal Supersymmetric Standard Model (MSSM)
\cite{HK85}, the situation is even more complicated.  First of all, due
to the richer particle content, there are more diagrams contributing to
the magnetic moment operators, and due to the larger number of free
parameters there are many potential additional sources of flavor
changing neutral currents \cite{BBM87}.  However, if one assumes further
that the MSSM is a low energy effective theory from minimal supergravity
\cite{Nil84} with radiative breaking of the electroweak symmetry, it is
known \cite{BBMR91,Bor93,Osh93,LNP93,Dia93,BG93,Oka93,GN93,HKT94,BV94}
that, besides the SM contribution mediated by the W, there are only
significant contributions from the charged Higgs ($\cHiggs$) and from
the chargino ($\chi^+_{1,2}$) exchange diagrams, while already with the
present experimental lower limits on the supersymmetry (SUSY) spectrum
the gluino contribution is small and the neutralino contribution is
always negligible.%
\footnote{This assumption is supported by present experimental data on
$\mathrm{B} \to X_s \gamma$ as well as the lack of evidence for large
contributions beyond the SM to other FCNC processes, e.g.\
$\bar\mathrm{K}^0 \mathrm{K}^0$ mixing and rare K decays.  For a
discussion of a scenario with a light gluino in the mass range 2--5~GeV
see e.g.~\cite{BR94} (and references quoted therein).}
The W and the $\cHiggs$ contributions always have the same sign in the
MSSM, but the chargino contribution can have either sign and may e.g.\
cancel the $\cHiggs$ contribution or (for small chargino masses and
large $\tan\beta$) even dominate the amplitude.

Since SUSY has to be broken, the mass splitting between the various
particles running in the loop can be very large, leading to additional
important QCD corrections.  We will advocate that, in a parameter space
analysis in the MSSM, one should not simply add up the contributions of
all diagrams at the W scale and use the renormalization group evolution
to run down this sum to the $b$ scale, but rather consider the
individual contributions separately.  This is especially important for
the chargino contribution, since the lightest chargino can be
significantly lighter than the W.

For the reasons mentioned above, we will ignore the contributions from
diagrams with gluinos and neutralinos in the present work.  They may
easily be included; corrections of the type considered in this work will
however always be numerically unimportant.

Our strategy will be similar to the work by Cho and Grinstein
\cite{CG91}.  Starting from the full theory at sufficiently high scales,
we will construct a series of effective theories that is well suited for
the description of the low-energy physics of interest.  We shall give
all ingredients that are necessary to obtain the leading QCD corrections
to the $\btosg$ inclusive rate and discuss some simple estimates for an
MSSM scenario with the assumptions mentioned above.  A full parameter
space analysis, which depends on the details of the implementation of
the soft SUSY-breaking, is however beyond the scope of the present paper
and will be discussed elsewhere.

This paper is organized as follows: In section 2 we briefly review the
elements of effective field theories needed for the present work.
Section 3 explains in detail the calculation for a type-II two-Higgs
doublet model, which is contained in the MSSM, while the contributions
of SUSY particles are discussed in section 4.  We shall present our
results in section 5 and finally conclude.

%%%%%%%%%%%%%%%%%%%%%%%%%%%%%%%%%%%%%%%%%%%%%%%%%%%%%%%%%%%%%%%%%%%%%%%%

\section{Effective field theory and $\btosg$}

The basic idea of effective field theories is by now well established,
and many excellent reviews have appeared in the literature
\cite{Wit77,Wei79,CGG84,Geo91,Geo93}.  Starting from some underlying
full theory, one integrates out the heavy degrees of freedom, thereby
producing a tower of non-renormalizable interactions (with couplings
proportional to inverse powers of the heavy particle mass) which contain
the virtual heavy particle effects.  One then runs the resulting
effective field theory down to the appropriate scale of interest using
the renormalization group.  If additional heavy particle thresholds are
crossed during the renormalization group running, then these particles
will also be integrated out.  The major advantages of using an effective
theory for the calculation of low-energy observables are convenience,
since calculations are usually simpler than in the full theory, and the
gain of insight.

A nontrivial feature of the effective field theory framework is the
automatic summation of large logarithms that originate from
perturbatively calculable short-distance physics by the renormalization
group.  As explained in detail e.g.\ in \cite{Geo93}, the
renormalization scale $\mu$ in a dimensional scheme (e.g.\
$\overline{\mathrm{MS}}$) serves to separate short-distance from
long-distance physics.  The Appelquist--Carazzone decoupling theorem
\cite{AC75} can be implemented properly in the (mass-independent)
$\overline{\mathrm{MS}}$ scheme by hand by matching the effective
theories below and above thresholds.  The advantage of having a
mass-independent scheme is that the renormalization group
$\beta$-functions do not explicitly depend on the scale $\mu$, while the
validity of the decoupling theorem guarantees that all intuitive
reasoning based on a so-called physical renormalization scheme remains
still true.

When the effective theory contains two heavy mass scales $m_1, m_2$ of
comparable magnitude, it is usually a good approximation to integrate
out both particles at a common scale.  On the other side, if the ratio
$x \equiv (m_1/m_2)^2$ is very small (i.e.\ $x \ll 1$), even if the
coupling constant is small, the product $\alpha \ln x$ may become of
order unity, and one is then forced to sum all powers of this product,
while corrections to the sum are suppressed by powers of $\alpha$ or
$x$.  Sometimes the situation is less favorable and lies somewhere in
between, as is the case e.g.\ for the SM with a heavy top quark
\cite{CG91} of, say, 175~GeV.  For the process $\btosg$, the most
important correction is the QCD running between the W and the $b$ scale,
whose size is (parametrically) given by
\[
  \alpha_S(\mW) \cdot \ln(\mW/\mb)^2 \simeq 0.7 \, ,
\]
while
\[
  \alpha_S(\mW) \cdot \ln(\mt/\mW)^2 \simeq (\mW/\mt)^2 \simeq 0.2
\]
indicates that one might miss numerically important pieces if either of
the latter would be neglected; compared to next-to-leading order
corrections which are of order $\alpha_S \simeq 0.1$.%
\footnote{Of course a full calculation of the next-to-leading order
corrections is necessary to resolve the well-known ambiguity in the
choice of scales in leading-order calculations.  This would require
however the computation of three-loop anomalous dimensions for the
process under consideration.}
What one can achieve with reasonable effort is to take into account the
resummation of the leading terms in the limit of a heavy top quark, and
then simply adding in the nonleading terms, thus neglecting terms which
are (up to logarithms) $O(\alpha (\mW/\mt)^2)$.  The choice of scale
where to add these nonleading terms is at this stage completely
arbitrary and can only be answered by a calculation of the power
corrections.  The remaining uncertainty is, however, less important than
neglected next-to-leading order corrections.

Let us now turn to the application to the $\btosg$ transition.  The
effective Hamiltonian of interest may be written as a sum of $\Delta
B=1$, $\Delta S=1$ operators:
\begin{equation}
\label{eq:H_eff}
 H_\mathrm{eff} =
 \frac{4 G_\mathrm{F}}{\sqrt 2} K_\mathrm{tb} K_\mathrm{ts}^*
 \sum_i C_i(\mu) {\cal O}_i(\mu)
\end{equation}
A suitable operator basis $\{ {\cal O}_i \}$ will be given below.

In general the definition of the operators in (\ref{eq:H_eff}) will
require the specification of a renormalization scheme.  From the fact
that the effective Hamiltonian is independent of the renormalization
scale, one derives renormalization group equations for the composite
operators ${\cal O}_i(\mu)$ and the coefficient functions $C_i(\mu)$.
The renormalization of a composite operator is formally defined in terms
of the divergent renormalization constants $Z_{ij}$ which relate
renormalized and bare operators:
\begin{equation}
{\cal O}_i^{\rm bare} = Z_{ij}(\mu) {\cal O}_j(\mu)
\end{equation}
Since the bare operators are $\mu$-independent, the renormalized
operators depend on the subtraction scale via the $\mu$ dependence of
the $Z_{ij}$:
\begin{equation}
\label{eq:RGE:O}
\mu \frac{d}{d\mu} {\cal O}_i =
\left( \mu \frac{d}{d\mu} Z^{-1}_{ij} \right) {\cal O}_j^{\rm bare} =
- \gamma_{ik} {\cal O}_k,
\end{equation}
where
\begin{equation}
\gamma_{ik} = Z^{-1}_{ij} \mu \frac{d}{d\mu} Z_{jk}
\end{equation}
is the so called anomalous dimension matrix.

{}From the scale independence of the effective Hamiltonian
(\ref{eq:H_eff}) one derives the renormalization group equations for the
Wilson coefficients $C_i$:
\begin{equation}
\label{eq:RGE:C}
\mu \frac{d}{d\mu} C_i(\mu) = \sum_j \left( \gamma^T \right)_{ij} C_j(\mu)
\end{equation}
If QCD corrections are neglected, the solution to this differential
equation is straightforward.  When QCD corrections are included, it
turns out to be favorable to eliminate the derivative with respect to
the renormalization scale in favor of a derivative with respect to the
coupling constant:
\begin{equation}
\label{eq:RGE:C-mod}
\beta \frac{d C_i}{d g_3} = \sum_j \left( \gamma^T \right)_{ij} C_j
\end{equation}
Here (and in the following) $g_3$ denotes the QCD coupling constant, and
$\beta = \mu (dg_3/d\mu)$ is the QCD beta function.

The solution to the differential equation (\ref{eq:RGE:C-mod}) is then
given by
\begin{eqnarray}
C(\mu) = T_{g} \left[ \exp\int\limits_{g_3(\mu_0)}^{g_3(\mu)} dg \,
\frac{\gamma^T(g)}{\beta(g)} \right] C(\mu_0),
\end{eqnarray}
where $T_{g}$ means an ordering in the coupling such that $g$ increases
from right to left (for $\mu<\mu_0$).  Since our anomalous dimension
matrices will be $g_3^2$ times a purely numerical matrix,
\[
\gamma =
 \gfrac \hat\gamma + O(g_3^4)
\]
the $g$-ordering is superfluous, and the $g$-integration is trivial:
\begin{equation}
\label{eq:RGE-solution}
C(\mu) =
 \exp\left[
 \left( \frac{1}{2b} \ln\frac{\alpha_3(\mu)}{\alpha_3(\mu_0)} \right)
 \hat\gamma^T \right] C(\mu_0)
\end{equation}

The most convenient way to calculate the anomalous dimension matrix
$\gamma$ is to consider Green functions with insertions of composite
operators.  Denote by $\Gamma^{(n)}_{{\cal O}_i}$ a renormalized
$n$-point 1PI Green function with one insertion of the operator ${\cal
O}_i$.  The anomalous dimension $\gamma_{ij}$ that determines the mixing
of ${\cal O}_i$ into ${\cal O}_j$ may then be simply read off from the
renormalization group equation for $\Gamma^{(n)}_{{\cal O}_i}$,
\begin{equation}
\gamma_{ij} \Gamma^{(n)}_{{\cal O}_j} =
- \left( \mu \frac{\partial}{\partial\mu} + \beta
\frac{\partial}{\partial g} + \gamma_m m \frac{\partial}{\partial m}
- n \gamma_{ext} \right) \Gamma^{(n)}_{{\cal O}_i}
\end{equation}
Here $\gamma_m = (\mu/m) (dm/d\mu)$, and $n\gamma_{\rm ext}$ accounts
for the wave-function anomalous dimensions arising from radiative
corrections to the external lines of the Green functions.

We shall use dimensional regularization with minimal subtraction
($\overline{\mathrm{MS}}$), $d = 4 - 2\eps$.  The
$\mathrm{SU}(3)_\mathrm{C} \times \mathrm{U}(1)_\mathrm{em}$ covariant
derivative then reads:
\begin{equation}
  D_\mu =
  \partial_\mu - i \mu^\eps g_3 G^a_\mu X^a - i \mu^\eps e A_\mu Q
\end{equation}
We will use the background field $R_\xi$ gauge \cite{Abb81} throughout
this work.  The anomalous dimensions of the fields are in this case
given by:
\begin{eqnarray}
&&
\gamma_\mathrm{quark} = \frac{2}{3} \gfrac, \quad
\gamma_\mathrm{squark} = -\frac{4}{3} \gfrac, \quad
\gamma_\mathrm{gluon} = \frac{\beta}{g_3},
\nonumber \\ &&
\gamma_\mathrm{m} = -4 \gfrac, \quad \;
\gamma_{\tilde\mathrm{m}} = -6 \gfrac, \qquad
\beta = b \frac{g_3^3}{8\pi^2} .
\end{eqnarray}
The coefficient appearing in the $\beta$-function has the value
\begin{equation}
b = - \frac{11}{2} + \frac{1}{3} n_f
    + \frac{1}{12} n_{\tilde q} + n_{\tilde g},
\end{equation}
where $n_f, n_{\tilde q}$ and $n_{\tilde g}$ are the number of active
quark flavors, squarks and gluinos, respectively.

%%%%%%%%%%%%%%%%%%%%%%%%%%%%%%%%%%%%%%%%%%%%%%%%%%%%%%%%%%%%%%%%%%%%%%%%

In all cases considered below, the operator basis of choice will contain
the following set of operators involving only light degrees of freedom
(i.e.\ photons, gluons, and ``light'' quarks with masses below $\mW$):%
\footnote{Note that our normalization differs from ref.~\cite{CG91}.
We have omitted their operator $\Pthree$, since the corresponding Wilson
coefficients will always be zero, and none of the operators under
consideration will mix back into it.}
\begin{eqnarray}
\noalign{\hbox{Dimension $d+1$:}}
\Oone   & = & - \loopfac \mb \bar s_\L D^2 b_\R
\nonumber \\
\label{eq:LE-Basis}
\Otwo   & = & \mu^\eps \mynorm{g_3} \mb
 \bar s_\L \sigma^{\mu\nu} X^a b_\R \; G^a_{\mu\nu}
\nonumber \\
\Othree & = & \mu^\eps \mynorm{e \Qb} \mb
 \bar s_\L \sigma^{\mu\nu} b_\R \; F_{\mu\nu}
\nonumber \\
\noalign{\hbox{Dimension $d+2$:}}
P^{1,A}_\L & = & - \mynorm{i}
 \bar s_\L T^A_{\mu\nu\sigma} D^\mu D^\nu D^\sigma b_\L
\nonumber \\
\Ptwo   & = & \mu^\eps \mynorm{e \Qb}
 \bar s_\L \gamma^\mu b_\L \; \partial^\nu F_{\mu\nu}
\nonumber \\
% \Pthree & = & \mu^\eps \mynorm{e \Qb}
%  F_{\mu\nu} \bar s_\L \gamma^\mu D^\nu b_\L
% \nonumber \\
\Pfour  & = & i \mu^\eps \mynorm{e \Qb}
 \bar s_\L \gamma^\mu \gamma^5 D^\nu b_\L \; \tilde F_{\mu\nu}
\end{eqnarray}
The tensors $T^A_{\mu\nu\sigma}$ appearing in $P_\L^{1,A}$, $A=1,\ldots
4$, are defined by:
\begin{eqnarray}
T^1_{\mu\nu\sigma} = g_{\mu\nu} \gamma_\sigma,
&&
T^2_{\mu\nu\sigma} = g_{\mu\sigma} \gamma_\nu,
\nonumber \\
T^3_{\mu\nu\sigma} = g_{\nu\sigma} \gamma_\mu,
&&
T^4_{\mu\nu\sigma} = -i \eps_{\mu\nu\sigma\tau} \gamma^\tau \gamma_5
\end{eqnarray}

In order to apply the procedure outlined above to the MSSM case, we will
consider in a first step the extension of the calculation by Cho and
Grinstein \cite{CG91} to the case of a type-II two-Higgs doublet model.
There are already two cases to consider, namely that the charged Higgs
can be either much lighter or much heavier than the $t$ quark.  We shall
then explain in detail how one adds to this picture the contributions
induced by the chargino loops.

%%%%%%%%%%%%%%%%%%%%%%%%%%%%%%%%%%%%%%%%%%%%%%%%%%%%%%%%%%%%%%%%%%%%%%%%

\section{Two-Higgs doublet model}

\subsection{$\mt > \mH$}

If the top quark is heavier than the W and the charged Higgs, then the
first step is to integrate out the top quark at the scale $\mu = \mt$.
This leads to an effective field theory for $\mu < \mt$ without the $t$,
but with new vertices of dimension larger than four that contain the
virtual $t$ effects.  For the process under consideration, we need, in
addition to the operator basis (\ref{eq:LE-Basis}), further operators.

In general, in the range $\mt > \mu > \mW$, one has to consider higher
dimensional operators which contain the W's, the \wbG/s $\phi_\pm$ and
the charged Higgs field.  By naive dimensional analysis, we expect that
higher dimensional operators are suppressed by inverse powers of the
ratio $\xtw \equiv (\mt/\mW)^2$.  Since this ratio is not very large for
phenomenologically acceptable top quark masses, the effects of the
higher dimensional operators are not necessarily small, compared to the
leading dimension 5 and dimension 6 operators.  Also, the matching
conditions at threshold in general are combinations of rational
functions and polynomials in $\xtw$.

Nevertheless, we shall take the approach motivated in the previous
chapter and keep only the leading operators and the leading terms in the
matching contributions.  Although we are unable to calculate the power
corrections, we shall later add the subleading terms in $1/\xtw$, so
that we get the same result in the limit of neglecting strong
corrections for $\mu > \mW$, as when all heavy particles are integrated
out simultaneously at the W scale.

The relevant part of the interaction Lagrangian in the charged current
sector reads
\begin{eqnarray}
\label{eq:W-int}
 L_\mathrm{CC} & = &
 \frac{g_2}{\sqrt 2} \, W^+_\mu \, \bar U \gamma^\mu K P_\L D +
 \mathrm{h.c.}
 \\ &+&
 \frac{g_2}{\sqrt 2 \mW}
 \left( \phi_+ \bar U \left[ M_U K P_\L - K M_D P_\R \right] D +
 \mathrm{h.c.} \right)
\nonumber
\end{eqnarray}
where $U = (u,c,t)$ and $D = (d,s,b)$ represent up-type and down-type
quarks, respectively, $M_U = \diag(m_\mathrm{u},m_\mathrm{c},\mt)$, $M_D
= \diag(m_\mathrm{d},m_\mathrm{s},\mb)$ are the quark mass matrices,
$g_2=e/\sin\theta_\mathrm{W}$ is the gauge coupling of
SU(2)${}_\mathrm{W}$, $P_{\L,\R}$ are projectors on the left- and
right-handed components of the fermions, and $K$ is the
Kobayashi-Maskawa matrix.  In the present work, we shall neglect the
masses of the quarks of the first two generations whenever appropriate.

{}From these expression one can see that the leading terms for $\xtw \gg
1$ come from vertices which involve the charged \wbG/s $\phi_\pm$ and
the top quark, since they are proportional to the top quark mass.  For
this reason, in the range $\mt > \mu > \mW$, we shall need, analogous to
the findings in \cite{CG91}, the following operators with external
\wbG/s, in addition to the operator basis (\ref{eq:LE-Basis}):
\begin{eqnarray}
Q_\LR & = & \mu^{2\eps} \mynorm{g_3^2} \mb
 \phi_+ \phi_- \, \bar s_\L b_\R
\nonumber \\
\Rone & = & i \mu^{2\eps} \mynorm{g_3^2}
 \phi_+ \phi_- \, \bar s_\L \Slash D b_\R
\nonumber \\
\Rtwo & = & i \mu^{2\eps} \mynorm{g_3^2}
 (D^\mu \phi_+) \phi_- \, \bar s_\L \gamma_\mu b_\R
 \\
\Rthree & = & i \mu^{2\eps} \mynorm{g_3^2}
 \phi_+ (D^\mu \phi_-) \bar s_\L \gamma_\mu b_\R
\nonumber
\end{eqnarray}
The inclusion of explicit factors $g_3^2$ into these operators is
motivated by the Gilman-Wise trick \cite{GW80}, which allows us to have
all one-loop contributions to the anomalous dimension matrices to be of
$O(g_3^2)$, so that the diagonalization of these matrices is scale
independent.  We will freely use this trick later on.

The interaction Lagrangian for the charged Higgs with the quarks reads:
\begin{equation}
\label{eq:H-int}
 L_{\cHiggs f\bar f'} = \frac{g_2}{\sqrt 2 \mW}
 \left( H^+ \bar U \left[ \cot\beta M_U K P_\L + \tan\beta K M_D P_\R
 \right] D + \mathrm{h.c.} \right),
 \end{equation}
with $\tan\beta = v_1/v_2$ being the ratio of the vacuum expectation
values of the Higgs fields which give rise to the masses of up- and
down-type quarks, respectively.

The interaction (\ref{eq:H-int}) has the same structure and quark mass
dependence of the couplings as the interaction of the would-be Goldstone
bosons $\phi_\pm$, see (\ref{eq:W-int}).  In the limit $\xth \equiv
(\mt/\mH)^2 \gg 1$, keeping only the leading terms in $1/\xth$, we are
lead to the following operators with charged Higgs bosons we have to add
to our operator basis in the range $\mt > \mu >
\mH$:
\begin{eqnarray}
Q_\LR'  & = & \mu^{2\eps} \mynorm{g_3^2} \mb H^+ H^- \, \bar s_\L b_\R
\nonumber \\
\Rone' & = & i \mu^{2\eps} \mynorm{g_3^2}
 H^+ H^- \, \bar s_\L \Slash D b_\R
\nonumber \\
\Rtwo' & = & i \mu^{2\eps} \mynorm{g_3^2}
 (D^\mu H^+) H^- \, \bar s_\L \gamma_\mu b_\R
 \\
\Rthree' & = & i \mu^{2\eps} \mynorm{g_3^2}
 H^+ (D^\mu H^-) \bar s_\L \gamma_\mu b_\R
\nonumber
\end{eqnarray}

% ----------------------------------------------------------------------

\subsubsection{Matching at $\mu = \mt$}

For $\mu > \mt$, our effective theory is a fully renormalizable theory,
which still contains all particles and interactions, so in this case all
coefficients of our effective Hamiltonian are zero:
\begin{equation}
C_i(\mu = \mt^+) = 0
\qquad \mbox{for all $i$}
\end{equation}
When we cross the $t$ threshold from above, i.e.\ when we integrate out
the top quark at $\mu = \mt$, we obtain the following changes to the
coefficients of the effective Hamiltonian, due the interactions from
the \wbG/s from matching the three-point functions $\Gamma^{bs\gamma}$
and $\Gamma^{bsg}$ \cite{CG91} (here $\Delta C_i = C_i(\mt^-) -
C_i(\mt^+)$):
\begin{eqnarray}
\label{Matching:mt:SM}
\DeltaphiW C_{\Oone} & = & - \myhalf
\nonumber \\
\DeltaphiW C_{\Otwo} & = &
 \frac{\Qb}{\Qt} \DeltaphiW C_{\Othree} = - \myhalf
\nonumber \\
\DeltaphiW C_{\Poneone} & = &
 \DeltaphiW C_{\Ponethree} = \frac{11}{18}
\nonumber \\
\DeltaphiW C_{\Ponetwo} & = & - \frac{8}{9}
\nonumber \\
\DeltaphiW C_{\Ponefour} & = &
 - \Qb \DeltaphiW C_{\Pfour} = \myhalf
\nonumber \\
\DeltaphiW C_{\Ptwo} & = & \frac{3}{4 \Qb}
\nonumber \\
\DeltaphiW C_{Q_\LR} & = & - \frac{16\pi^2}{g_3^2}
\nonumber \\
\DeltaphiW C_{\Rone} & = & \DeltaphiW C_{\Rtwo} = \frac{16\pi^2}{g_3^2}
\nonumber \\
\DeltaphiW C_{\Rthree} & = & 0
\end{eqnarray}
Similarly, there are contributions from the interactions with the
charged Higgs bosons:%
\footnote{We prefer to keep the contributions from each interaction
separately, for there are different cases below.}
\begin{eqnarray}
\label{Matching:mt:Higgs}
\DeltaH C_{\Oone} & = & \frac{1}{2}
\nonumber \\
\DeltaH C_{\Otwo} & = &
 \frac{\Qb}{\Qt} \DeltaH C_{\Othree} = \frac{1}{2}
\nonumber \\
\DeltaH C_{\Poneone} & = &
 \DeltaH C_{\Ponethree} = \frac{11}{18} \cot^2\beta
\nonumber \\
\DeltaH C_{\Ponetwo} & = & - \frac{8}{9} \cot^2\beta
\nonumber \\
\DeltaH C_{\Ponefour} & = &
 - \Qb \DeltaH C_{\Pfour} = \frac{1}{2} \cot^2\beta
\nonumber \\
\DeltaH C_{\Ptwo} & = & \frac{3}{4 \Qb} \cot^2\beta
\nonumber \\
\DeltaH C_{Q_\LR'}  & = & \frac{16\pi^2}{g_3^2}
\nonumber \\
\DeltaH C_{\Rone'} & = & \DeltaH C_{\Rtwo'} =
 \frac{16\pi^2}{g_3^2} \cot^2\beta
\nonumber \\
\DeltaH C_{\Rthree'} & = & 0
\end{eqnarray}
At this point it is worthwhile to note that, had we not matched at the
scale $\mu = \mt$ but at a different scale (or used a different
subtraction scheme), we would have found logarithmic contributions in
the matching corrections to the coefficient of $\Ptwo$:
\begin{equation}
\label{eq:matching-logs:t}
\DeltaphiW C_{\Ptwo} = \frac{1}{\Qb}
 \left[ \frac{3}{4} + \frac{1}{6} \ln\frac{\mu^2}{\mt^2} \right] ,
\quad
\DeltaH C_{\Ptwo} = \frac{1}{\Qb}
 \left[ \frac{3}{4} + \frac{1}{6} \ln\frac{\mu^2}{\mt^2} \right] \cot^2\beta
\end{equation}
These logarithms which vanish for $\mu=\mt$ are regenerated at lower
scales by the renormalization group for the effective theory below
$\mt$.  It is therefore not surprising that they are present in the full
expressions for this coefficient given in the appendix, when both
particles in a loop are integrated out at the same scale; there it
appears as an unsuppressed logarithm of the mass ratio of the particles
in the loop.

% ----------------------------------------------------------------------

\subsubsection{Running below $\mt$}

The anomalous dimension matrices for the mixing of the operators $O_i$
and $P_i$ has already been given in ref.~\cite{CG91}.  For completeness,
we quote the result obtained in this work.

First, there is a mixing of the operators $Q$, $R$ with \wbG/
fields into the operators without ($O$, $P$):
\begin{equation}
\label{ADM:SM:mixback}
\hat\gamma =
\bordermatrix{
        & O_\LR & P_\L^{1,A} & \Ptwo  & \Pfour \cr
Q_\LR   & 0     & 0          & 0      & 0 \cr
\Rone   & 0     & 0          & 0      & 0 \cr
\Rtwo   & 0     & 0          & 1/6\Qb & 0 \cr
\Rthree & 0     & 0          &-1/6\Qb & 0 \cr
}
\end{equation}
Note that this mixing back is of order $O(g_3^2)$ due to our choice of
the coefficients in front of the operators $Q$, $R$, and not due to
``proper'' QCD corrections.

For the QCD-induced entries in the anomalous dimension matrix, one has
\begin{equation}
\label{ADM:OP}
\hat\gamma = \bordermatrix{
      &\Oone&\Otwo&\Othree
      & \Poneone&\Ponetwo&\Ponethree&\Ponefour&\Ptwo&\Pfour \cr
\Oone & {20 \over 3} & 1 & -2 & 0 & 0 & 0 & 0 & 0 & 0 \cr
\Otwo & -8 & {2\over 3}& {4\over 3} & 0 & 0 & 0 & 0 & 0 & 0 \cr
\Othree & 0 & 0 & {16\over 3} & 0 & 0 & 0 & 0 & 0 & 0 \cr
\Poneone & 6 & 2 & -1 & {2\over 3} & 2 & -2 & -2 & 0 & 0 \cr
\Ponetwo & 4 & {3\over 2} & 0 & -{113\over 36} &{137\over 18} & -{113\over 36}
      & -{4\over 3} & {9\over 4} & 0 \cr
\Ponethree & 2 & 1 & 1 & -2 & 2 & {2 \over 3} & -2 & 0 & 0 \cr
\Ponefour& 0 & {1\over 2} & 2 & -{113\over 36} & {89\over 18} & -{113\over 36}
       & {4 \over 3} & {9\over 4} & 0 \cr
P^2_\L & 0 & 0 & 0 & 0 & 0 & 0 & 0 & 0 & 0 \cr
P^4_\L & 0 & 0 & {4\over 3} & 0 & 0 & 0 & 0
    & 0 & 0 \cr
}
\end{equation}
Similarly, the mixing among the operators with \wbG/ fields is known to be:
\begin{equation}
\label{ADM:SM:phi}
\hat\gamma =
\bordermatrix{
        & Q_\LR & \Rone  & \Rtwo & \Rthree \cr
Q_\LR   & -2b   & 0      & 0     & 0       \cr
\Rone   & 0     & -2b    & 0     & 0       \cr
\Rtwo   & 0     & 0      & -2b   & 0       \cr
\Rthree & 0     & 0      & 0     & -2b     \cr
}
\end{equation}
Obviously, the same mixing matrices are found when one considers the
mixing of the operators with charged Higgs fields, i.e.\ when one
replaces $Q_\LR \to Q_\LR'$, $R_\L^i \to R_\L^i{}'$ in
eqs.~(\ref{ADM:SM:mixback}) and (\ref{ADM:SM:phi}).

% ----------------------------------------------------------------------

\subsubsection{Matching at $\mu = \mH$ and $\mu = \mW$}
\label{sec:subleading}

In the process of scaling down, when we encounter the charged Higgs or W
threshold, we have to integrate out the $\cHiggs$ or W and \wbG/s,
respectively.  Due to decoupling that has to take place below threshold,
we shall remove the operators $Q'$, $R'$ from our operator basis for
$\mu < \mH$ and $Q$, $R$ for $\mu < \mW$.  Again we obtain the finite
changes of the coefficients of the operators $O$ and $P$ by matching
Green functions calculated in the theories above and below threshold.

Since we neglect small terms proportional to $m_\mathrm{u}$ or
$m_\mathrm{c}$, we find no nonvanishing contribution from the matching
of the effective theories above and below $\mu = \mH$, i.e.\ our Wilson
coefficients are continuous:
\begin{equation}
\label{Matching:mh:light-higgs}
C_i(\mH^+) = C_i(\mH^-)
\end{equation}

Matching the effective theories above and below $\mu = \mW$, we find the
following changes in the coefficients of the effective Hamiltonian (here
$\Delta C = C(\mW^-) - C(\mW^+)$):
\begin{eqnarray}
\label{Matching:mw:SM}
\DeltaphiW C_{\Oone} & = &
 \DeltaphiW C_{\Otwo} = \DeltaphiW C_{\Othree} = 0
\nonumber \\
\DeltaphiW C_{\Poneone} & = &
 \DeltaphiW C_{\Ponethree} = \frac{2}{9}
\nonumber \\
\DeltaphiW C_{\Ponetwo} & = & - \frac{7}{9}
\nonumber \\
\DeltaphiW C_{\Ponefour} & = & 1
\nonumber \\
\DeltaphiW C_{\Ptwo} & = & \frac{1}{2 \Qb}
\nonumber \\
\DeltaphiW C_{\Pfour} & = & - \frac{3}{\Qb}
\end{eqnarray}
Again, had we matched at a different scale $\mu \neq \mW$, we would have
found different matching contributions for some of the coefficients:
\begin{eqnarray}
\label{eq:matching-logs:W}
\DeltaphiW C_{\Poneone} & = &
 \DeltaphiW C_{\Ponethree} =
 \frac{2}{9} + \frac{2}{3} \ln\frac{\mu^2}{\mW^2}
\nonumber \\
\DeltaphiW C_{\Ponetwo} & = &
 -\frac{7}{9} - \frac{4}{3} \ln\frac{\mu^2}{\mW^2}
\nonumber \\
\DeltaphiW C_{\Ptwo} & = &
 \frac{1}{\Qb}
 \left( \myhalf + \frac{2}{3} \ln\frac{\mu^2}{\mW^2} \right)
\end{eqnarray}
But the coefficients of $\ln\mu^2$ are just the coefficients of those
logarithms in eqs.~(\ref{eq:1-loop:W}) which give the leading
(divergent) contribution to the $C_i$ in the limit of small quark
masses.  These logarithms are regenerated by the renormalization group
running in the low energy effective theory valid at scales $\mu < \mW$
and therefore need not be discussed here any further.

We shall to now use our freedom to add subleading terms in $1/\xtw$,
$1/\xth$ to the coefficients $C_i$.  In order to see how this is
accomplished, let us for the moment neglect the proper QCD corrections,
so we have to consider only the entries in the anomalous dimension
matrix given in (\ref{ADM:SM:mixback}).  Solving the renormalization
group equations (\ref{eq:RGE:C}), we find that only one coefficient runs
below $\mt$,
\begin{equation}
\label{eq:running:noQCD}
C_\Ptwo(\mu) = C_\Ptwo(\mt) +
  \left( \frac{1}{6 \Qb} + \frac{1}{6 \Qb} \cot^2\beta \right)
  \log\frac{\mu^2}{\mt^2}
\end{equation}
where the first term in parentheses is due to the mixing of $R_\L^2$
into $\Ptwo$, and the second due to $R_\L^2{}'$.  We see that the
renormalization group reproduces the logarithmic terms already discussed
in eq.~(\ref{eq:matching-logs:t}), which would have been there, had we
done the matching at a different scale.

The subleading contributions are found by taking the standard one-loop
result from integrating out both particles in the loop at the same
scale, see appendix, and subtracting the leading contributions that we
have found from the matching contributions
(\ref{Matching:mt:SM},\ref{Matching:mt:Higgs}) and the running
(\ref{eq:running:noQCD}) without QCD.  We will always refer to this
procedure for obtaining the subleading terms in the rest of the present
work.

Let us for the moment assume that $\mH > \mW$.  When we integrate out
the charged Higgs at $\mu=\mH$, we obtain the subleading contributions
from (\ref{Matching:mt:Higgs},\ref{eq:running:noQCD},\ref{eq:1-loop:H}):
\begin{eqnarray}
\label{Matching:mh:subleading}
\DeltaHsub C_{\Oone} & = & \xth F_4(\xth) - \myhalf
\nonumber \\
\DeltaHsub C_{\Otwo} & = & \frac{\Qb}{\Qt} \DeltaHsub C_{\Othree} =
 \frac{\xth}{2} (F_3(\xth) + F_4(\xth)) - \myhalf
\nonumber \\
\DeltaHsub C_{\Poneone} & = & \DeltaHsub C_{\Ponethree}
\nonumber \\ & = &
 \left(
 \frac{\xth}{3}\left( 2 F_2(\xth) + F_3(\xth) + 2 F_4(\xth) \right)
 - \frac{11}{18} \right) \cot^2\beta
\nonumber \\
\DeltaHsub C_{\Ponetwo} & = &
 \left(
 \frac{2\xth}{3} \left( F_2(\xth) - F_3(\xth) - 2 F_4(\xth) \right)
 + \frac{8}{9} \right) \cot^2\beta
\nonumber \\
\DeltaHsub C_{\Ponefour} & = & - \Qb \DeltaHsub C_{\Pfour} =
 \left( \xth F_4(\xth) - \myhalf \right) \cot^2\beta
 \\
\DeltaHsub C_{\Ptwo} & = &
 \frac{1}{\Qb} \left(
 \xth \left( \myhalf F_3(\xth) + F_4(\xth) \right)
 - \frac{3}{4} - \frac{\ln\xth}{6(\xth-1)} \right) \cot^2\beta
\nonumber
\end{eqnarray}
The functions $F_i(x)$ are given in appendix A.  One may easily verify
that the terms on the r.h.s.\ are of order $O(1/\xth)$, so they are
truly subleading.  Especially there is no (leading) logarithmic
dependence of the matching contributions to $C_\Ptwo$ on the mass ratio
$\xth$, since all such dependencies must come from the renormalization
group.

Of course there is an ambiguity in the choice of scale where to add the
subleading contributions.  This ambiguity can only resolved by computing
the power corrections, which fortunately differ from our naively adding
the subleading terms only by a next-to-leading contribution.  We shall
define our procedure by assuming that taking the scale equal to the mass
of the lightest particle in the loop is a suitable choice.

After scaling down from $\mu=\mH$ and adding in the leading matching
contributions at $\mu=\mW$, we will also consider the subleading
contributions.  In analogy to the previous case we find:
\begin{eqnarray}
\label{Matching:mw:subleading}
\DeltaphiWsub C_{\Oone} & = &
 -\xtw F_4(\xtw) + \myhalf
\nonumber \\
\DeltaphiWsub C_{\Otwo} & = & \frac{\Qb}{\Qt} \DeltaphiWsub C_{\Othree} =
 -\frac{\xtw}{2} \left(F_3(\xtw) + F_4(\xtw) + \myhalf \right)
\nonumber \\
\DeltaphiWsub C_{\Poneone} & = & \DeltaphiWsub C_{\Ponethree}
\nonumber \\ & = &
 \frac{\xtw+2}{3} \left(2 F_2(\xtw) + F_3(\xtw) + 2 F_4(\xtw) \right)
  - \frac{11}{18}
\nonumber \\
\DeltaphiWsub C_{\Ponetwo} & = &
 \frac{2(\xtw+2)}{3} \left(F_2(\xtw) - F_3(\xtw) - 2 F_4(\xtw) \right)
  + \frac{8}{9}
\nonumber \\
\DeltaphiWsub C_{\Ponefour} & = &
 (\xtw-2) F_4(\xtw) - \myhalf
\nonumber \\
\DeltaphiWsub C_{\Ptwo} & = &
 \frac{1}{\Qb}
 \left( (\xtw+2) \left( \myhalf F_3(\xtw) + F_4(\xtw) \right)
  - \frac{\ln\xtw}{2(x-1)} - \frac{3}{4} \right)
\nonumber \\
\DeltaphiWsub C_{\Pfour} & = &
 \frac{1}{\Qb}
 \left( \frac{7}{2} - 2\xth F_3(\xth) - 5\xth F_4(\xth) \right)
\end{eqnarray}

% ----------------------------------------------------------------------

\subsubsection{Reduction by equations of motion}

In order to be able to use the results from previous calculations for
the running between the W and the $b$ scale, we have to match our
operator basis to the operator basis employed there.  To this end, we
use the equations of motions, as in ref.\ \cite{CG91}.  For the
effective Hamiltonian just below the W scale, one then finds:
\begin{eqnarray}
\label{eq:EOM}
 H_{\mathrm{eff}}  & = &
 \frac{4 G_\mathrm{F}}{\sqrt 2} K_\mathrm{tb} K_\mathrm{ts}^*
 \sum_i C_i(\mW^-) O_i(\mW^-)
\\
& \stackrel{{\rm EOM}}{\to} &
 \frac{4 G_\mathrm{F}}{\sqrt 2} K_\mathrm{tb} K_\mathrm{ts}^*
\left[
\left(
-\myhalf C_{\Oone}
+      C_{\Otwo}
-\myhalf C_{\Poneone}
-\quarter C_{\Ponetwo}
+\quarter C_{\Ponefour}
\right)
\Otwo
\right. \nonumber \\ && \left.
+ \left(
-\myhalf C_{\Oone}
+      C_{\Othree}
-\myhalf C_{\Poneone}
-\quarter C_{\Ponetwo}
+\quarter C_{\Ponefour}
-\quarter C_{\Pfour}
\right)
\Othree
\right]
\nonumber
\\ && + \; \frac{g_3^2}{16\pi^2} \; \mbox{(four-fermion operators)}
 \nonumber
\end{eqnarray}
Since we are only interested in the leading contributions from the QCD
corrections due to a large mass splitting, we may drop the contributions
to the four-fermion operators in (\ref{eq:EOM}) since these are
suppressed by a factor $g_3^2/16\pi^2$ and therefore truly nonleading.

To this expression we have to add of course the standard four-fermion
operators $(\bar b_L \gamma_\mu s_L) (\bar q_L \gamma^\mu q_L)$, $q=u,c$
(with the appropriate CKM mixing coefficients) from integrating out the
W.  The Wilson coefficients obtained this way at the W scale may then be
used as input for the renormalization group running
\cite{CCRV90,CFMRS93,BMMP93} down to the $b$ scale.

%%%%%%%%%%%%%%%%%%%%%%%%%%%%%%%%%%%%%%%%%%%%%%%%%%%%%%%%%%%%%%%%%%%%%%%%

\subsection{$\mH > \mt$}

If the charged Higgs is heavier than the top quark, the picture becomes
a little more involved.  As we run down from large scales, we first
encounter the threshold of the charged Higgs.  Therefore, as a first
step, we integrate out the charged Higgs.  In the same spirit as in the
previous case, we shall now be mainly concerned with the leading
contributions in the limit $\xth \ll 1$.

In the range $\mH > \mu > \mt$, after integrating out the charged Higgs,
we have to deal with four-fermion operators of dimension~6 that involve
a $b$, an $s$, plus a quark--anti-quark pair.  Besides the operators
(\ref{eq:LE-Basis}), our operator basis contains:
\begin{eqnarray}
\label{eq:Operators:S}
S_1 & = &
 (\bar s_\L^\alpha \gamma_\mu b_\L^\alpha)
 (\bar t_\R^\beta  \gamma^\mu t_\R^\beta)
\nonumber \\
S_2 & = &
 (\bar s_\L^\alpha \gamma_\mu b_\L^\beta)
 (\bar t_\R^\beta  \gamma^\mu t_\R^\alpha)
\nonumber \\
S_3 & = &
(\bar s_\L^\alpha \gamma_\mu b_\L^\alpha)
\sum_q \left(\bar q_\L^\beta  \gamma^\mu q_\L^\beta \right)
\nonumber \\
S_4 & = &
(\bar s_\L^\alpha \gamma_\mu b_\L^\beta)
\sum_q \left(\bar q_\L^\beta  \gamma^\mu q_\L^\alpha \right)
\nonumber \\
S_5 & = &
(\bar s_\L^\alpha \gamma_\mu b_\L^\alpha)
\sum_q \left(\bar q_\R^\beta  \gamma^\mu q_\R^\beta \right)
\nonumber \\
S_6 & = &
(\bar s_\L^\alpha \gamma_\mu b_\L^\beta)
\sum_q \left(\bar q_\R^\beta  \gamma^\mu q_\R^\alpha \right)
\nonumber \\
S_7 & = & \mu^{2\eps} \mynorm{g_3^2} \frac{\mb}{\mt}
(\bar s_\L^\alpha t_\R^\beta) (\bar t_\L^\beta b_\R^\alpha)
\nonumber \\
S_8 & = & \mu^{2\eps} \mynorm{g_3^2} \frac{\mb}{\mt}
(\bar s_\L^\alpha t_\R^\alpha) (\bar t_\L^\beta b_\R^\beta)
\nonumber \\
S_9 & = & \quarter \mu^{2\eps} \mynorm{g_3^2} \frac{\mb}{\mt}
(\bar s_\L^\alpha \sigma_{\mu\nu} t_\R^\beta)
(\bar t_\L^\beta  \sigma^{\mu\nu} b_\R^\alpha)
\nonumber \\
S_{10} & = & \quarter \mu^{2\eps} \mynorm{g_3^2} \frac{\mb}{\mt}
(\bar s_\L^\alpha \sigma_{\mu\nu} t_\R^\alpha)
(\bar t_\L^\beta  \sigma^{\mu\nu} b_\R^\beta)
\end{eqnarray}
Here $\alpha,\beta$ are color indices of the quarks, and the sums run
over all active flavors.  Again, the inclusion of the additional factors
$g_3^2$ is motivated by the Gilman-Wise trick \cite{GW80}, as are the
factors $\mb/\mt$ to keep the anomalous dimension matrices mass
independent.  The different normalization of $S_1 \ldots S_6$ and $S_7
\ldots S_{10}$ will be explained below.

% \subsubsection{Matching at $\mu=\mH$}

Integrating out the charged Higgs at $\mu = \mH$, we find to leading
order in $\xth \equiv (\mt/\mH)^2$:
\begin{eqnarray}
\label{Matching:heavy-H:mh}
C_{\Oone} & = & \frac{1}{2} \xth
\nonumber \\
C_{\Otwo} & = & \frac{\Qb}{\Qt} C_{\Othree} =  -\frac{1}{2} \xth
\nonumber \\
C_{\Poneone} & = & C_{\Ponethree} = -\frac{1}{9} \xth \cdot \cot^2\beta
\nonumber \\
C_{\Ponetwo} & = & \frac{7}{18} \xth \cdot \cot^2\beta
\nonumber \\
C_{\Ponefour} & = & - \Qb C_{\Pfour} =
 \frac{1}{2} \xth \cdot \cot^2\beta
\nonumber \\
C_{\Ptwo} & = &
 \frac{1}{\Qb} \left( -\frac{1}{4} \xth \right) \cot^2\beta
\nonumber \\
C_{S_2} & = & -\myhalf \xth \cot^2 \beta
\nonumber \\
C_{S_8} & = & \frac{16\pi^2}{g_3^2} \xth
\nonumber \\
C_{S_i} & = & 0, \qquad i = 1, 3 \ldots 7, 9, 10
\end{eqnarray}

Let us start again with the mixing back of the operators $S$ into the
operators $O$ and $P$.  Because of the chirality structure of the
operators, we find two different situations at one loop.  The operators
$S_1, \ldots S_6$ appear to have a zeroth order mixing ($g_3^0$) at one
loop into the operators $P$.
\begin{equation}
\label{mixing:P:S1-S6}
\gamma^{(0)} =
\bordermatrix{
    & \Poneone       & \Ponetwo        & \Ponethree     & \Ponefour
     & \Ptwo          & \Pfour \cr
S_1 & 0              & 0               & 0              & 0
     & 2\QtQb         & 0 \cr
S_2 &  \frac{2}{3}   & -\frac{4}{3}  & \frac{2}{3}      & 0
     &  \frac{2}{3\Qb} & 0 \cr
S_3 & \frac{4}{3}    & -\frac{8}{3}    & \frac{4}{3}    & 0
     & \frac{2}{\Qb} \sum_q Q_q & 0 \cr
S_4 & \frac{2n_f}{3} & -\frac{4n_f}{3} & \frac{2n_f}{3} & 0
     & \frac{2}{3}\left(6+\frac{1}{\Qb} \sum_q Q_q -n_f \right) & 0 \cr
S_5 & 0              & 0               & 0              & 0
     & \frac{2}{\Qb} \sum_q Q_q & 0 \cr
S_6 & \frac{2n_f}{3} & -\frac{4n_f}{3} & \frac{2n_f}{3} & 0
     & \frac{2}{3}\left(\frac{1}{\Qb} \sum_q Q_q -n_f \right) & 0 \cr
}
\end{equation}
However, by inspection of the equations of motion (\ref{eq:EOM}) one
sees that the back mixing vanishes to this order; therefore we may
simply drop this contribution.  As is well known, one has to consider
this mixing at two-loop order.  The anomalous dimension matrix can be
derived from eq.~(25) of ref.~\cite{CFMRS93}, and reads in our
normalization
\begin{equation}
\label{mixing:O:S1-S6}
\hat\gamma =
\bordermatrix{
    & \Otwo                             & \Othree \cr
S_1 &               -\frac{3}{2}        & 0 \cr
S_2 &               -\frac{119}{54}     & \frac{224}{27} \cr
S_3 & \frac{70}{27} +\frac{3}{2} n_f    & \frac{232}{27} \cr
S_4 &   3 +          \frac{35}{27} n_f  & \frac{8}{27} n_f + 4 \bar n_f \cr
S_5 & -\frac{7}{3} - \frac{3}{2} n_f    & -\frac{16}{3} \cr
S_6 & - 2 -          \frac{119}{54} n_f & \frac{8}{27} n_f - 4 \bar n_f \cr
}
\end{equation}
Here $n_f = n_u + n_d$ is the number of active flavors, and $\bar n_f =
n_d + (Q_u/Q_d) n_u$.

On the other hand, the mixing of $S_7 \ldots S_{10}$ into the operators
$O$ does not vanish at one loop:
\begin{equation}
\label{mixing:O:S7-S10}
\hat\gamma =
\bordermatrix{
       & \Oone & \Otwo  & \Othree \cr
S_7    & 0     & 0      & -\frac{3}{2} \QtQb \cr
S_8    & 0     & -\myhalf & -\myhalf \QtQb       \cr
S_9    & 0     & 0      &  \frac{3}{2} \QtQb \cr
S_{10} & 0     &  \myhalf &  \myhalf \QtQb       \cr
}
\end{equation}
Again one may verify that these entries of the ADM are consistent with
the $\ln\mu$-dependence of the matching contributions
(\ref{Matching:heavy-H:mh}).

Let us now turn to the mixing among the four-fermion operators.  Since
the considered operators are all of dimension $d+2$, and because the QCD
interactions preserve chirality, the operators $S_{1,\ldots,6}$ and the
operators $S_{7,\ldots,10}$ will mix only among themselves,
respectively.

The one-loop mixing among the $S_1 \ldots S_6$ is well known \cite{GW80}:
\begin{equation}
\label{ADM:S1-S6}
\hat\gamma =
\bordermatrix{
    & S_1 & S_2 & S_3             & S_4
                 & S_5             & S_6 \cr
S_1 & 1   & -3  & 0               & 0               & 0    & 0 \cr
S_2 & 0   & -8  &  -\frac{1}{9}   & \frac{1}{3}
                 &  -\frac{1}{9}   & \frac{1}{3} \cr
S_3 & 0   & 0   &  -\frac{11}{9}  & \frac{11}{3}
                 &  -\frac{2}{9}   & \frac{2}{3} \cr
S_4 & 0   & 0   & 3-\frac{n_f}{9} & \frac{n_f}{3}-1
                 &  -\frac{n_f}{9} & \frac{n_f}{3} \cr
S_5 & 0   & 0   & 0               & 0               & 1    & -3 \cr
S_6 & 0   & 0   &  -\frac{n_f}{9} & \frac{n_f}{3}
                 &  -\frac{n_f}{9} & \frac{n_f}{3}-8 \cr
}
\end{equation}
For the mixing of $S_7 \ldots S_{10}$ we find:
\begin{equation}
\label{ADM:S7-S10}
\hat\gamma =
\bordermatrix{
       & S_7  & S_8   & S_9              & S_{10} \cr
S_7    & 1-2b & -3    & -\frac{7}{3}     & -1     \cr
S_8    & 0    & -8-2b & -2               & \frac{2}{3} \cr
S_9    & -7   & -3    & -\frac{19}{3}-2b & 0 \cr
S_{10} & -6   &  2    & 0                & \frac{8}{3}-2b \cr
}
\end{equation}
Since the operators $O_\LR$ are dimension $d+1$, there is no mixing back
into $S_{1,\ldots,10}$.

Note that with our chosen normalization of the operators as given in
(\ref{eq:Operators:S}) all relevant mixing occurs at order $g_3^2$, and
all entries in the anomalous dimension matrix are dimensionless.

% \subsubsection{Matching at $\mu=\mt$}

After running down to $\mu=\mt$, we integrate out the $t$ quark.  As far
as the operators $S_1,S_2,S_{7,\ldots,10}$ are concerned, they are just
removed, since they give no contribution to the matching; for the
operators $S_{3,\ldots,6}$ the $t$ quark has to be excluded from the
sum, because it is ``inactive'' for $\mu<\mt$.  Again we will take into
account the subleading terms in $\xth$ according to the general
prescription given in section \ref{sec:subleading}.  Then we will
continue as in the case for the Standard Model with a heavy top, except
that the coefficients $C_i(\mt^+)$ are now nonvanishing.

%%%%%%%%%%%%%%%%%%%%%%%%%%%%%%%%%%%%%%%%%%%%%%%%%%%%%%%%%%%%%%%%%%%%%%%%

\section{Supersymmetric contributions}

\subsection{Flavor changing chargino interactions}

Let us in analogy to \cite{HK85} denote by $\tilde W^\pm, \, \tilde
H_1^-$ and $\tilde H_2^+$ the superpartners of the W and the charged
components of the Higgs fields, respectively.  Define the two component
spinors $\psi_j^\pm$ by
\begin{equation}
 \psi_j^+ = \left( -i \tilde W^+, \tilde H_2^+ \right)
 , \quad
 \psi_j^- = \left( -i \tilde W^-, \tilde H_1^- \right)
 , \qquad j=1,2
\end{equation}
The mass term for the W-inos and higgsinos then takes the following
form:
\begin{equation}
 L_{\cal M} = -
 \psi^-  {\cal M} \psi^+
 + \mbox{h.c.}
\end{equation}
where the mass matrix is given by
\begin{equation}
\label{eq:mass-matrix}
 {\cal M} =
 \pmatrix{M_2 & \sqrt2 \mW \sin\beta \cr \sqrt2 \mW \cos\beta & \muh} .
\end{equation}
where $M_2$ is the soft SUSY breaking mass term for the W-inos at the
weak scale, and $\muh$ is the renormalized Higgs mixing parameter.

This mass matrix may be diagonalized with the help of two unitary
matrices $U,V$ such that
\begin{equation}
  U^* {\cal M} V^\dagger =
  {\cal M}_\chi = \diag(\tilde m_1, \tilde m_2)
\end{equation}
is a diagonal matrix with nonnegative entries.  The corresponding
charged mass-eigenstate 4-spinors are the charginos
\begin{equation}
\chi_i^+ = \pmatrix{V_{ij} \psi_j^+ \cr U^*_{ij} \bar\psi_j^-} .
\end{equation}
We shall find it however more convenient to rewrite the interactions of
the charginos by their charge conjugates
\begin{equation}
\label{def:charginos}
\chi_i^- \equiv \left(\chi_i^+ \right)^c = C (\bar\chi_i^+)^T
         = \pmatrix{U_{ij} \psi_j^- \cr V^*_{ij} \bar\psi_j^+} ,
\end{equation}
so whenever we refer to charginos below, we mean the $\chi_i^-$ given in
(\ref{def:charginos}).

Let us apply these definitions to the interactions of the charged
gauginos and higgsinos and convert to 4-spinor notation.  Neglecting for
the moment the mixing of quarks and of squarks and concentrating on the
terms involving $b$ quarks, the relevant Lagrangian for
chargino-quark-squark interactions reads:
\begin{equation}
 L_{\chi b \tilde t} =
- g_2 V^*_{i1} \tilde t_\L^\dagger (\bar\chi_i P_\L b)
+ g_2 \lambda_\mathrm{t} V^*_{i2} \tilde t_\R^\dagger (\bar\chi_i P_\L b)
+ g_2 \lambda_\mathrm{b} U_{i2}   \tilde t_\L^\dagger (\bar\chi_i P_\R b)
+ {\rm h.c.} \; ,
\end{equation}
and the couplings $\lambda_q$ are proportional to the Yukawa couplings:
\begin{equation}
\label{eq:Yukawa}
\lambda_\mathrm{t} = \frac{\mt}{\sqrt2 \mW \sin\beta}
\quad , \quad
\lambda_\mathrm{b} = \frac{\mb}{\sqrt2 \mW \cos\beta} \; .
\end{equation}
A similar expression is found for the interactions of the charginos with
the quarks and squarks of the second family.  In this case one can
however neglect the terms proportional to $\lambda_\mathrm{c}$,
$\lambda_\mathrm{s}$, which originate in the coupling of the higgsino
components of the charginos to the quark and squark fields.

Since the Yukawa couplings of the matter fields to the Higgs fields are
not flavor diagonal in a weak interaction basis, we have to take into
account the mixing among quarks and among squarks.  Let us denote as in
ref.\ \cite{BBMR91} by $\tilde q_{l\L,\R}$ the squark current
eigenstates (where $q=u,d$, and $l=1,2,3$ is the generation label),
$\tilde q_a$ ($a=1,\ldots,6$) the corresponding mass eigenstates with
masses $\tilde m_a$.  We define the $6\times3$ squark mixing matrices
$\Gamma_{Q\L,\R}$ by
\begin{equation}
\label{def:squark-mixing}
\tilde q_{\L,\R} = \Gamma_{Q\L,\R}^\dagger \tilde q \; .
\end{equation}
The relevant chargino interactions involving down-type quarks may then
be written as
\begin{equation}
 L_{\chi d \tilde u} =
-g_2 \sum_{j,a,l} \left[
 \tilde u_a^\dagger \;\bar\chi_j
 \left( G^{jal} P_\L - H^{jal} P_\R \right) d_l
 \right]
 + {\rm h.c.} \; ,
\end{equation}
where
\begin{eqnarray}
 G^{jal} & = & V_{j1}^*  \Gamma_{U\L}^{al}
             - V_{j2}^* (\Gamma_{U\R} \Lambda_U K)^{al}
 \nonumber \\
 H^{jal} & = & U_{j2}   (\Gamma_{U\L} \Lambda_D  )^{al}
\end{eqnarray}
Here $\Lambda_U = M_U/(\sqrt2 \mW \sin\beta)$,
$\Lambda_D = M_D/(\sqrt2 \mW \cos\beta)$, are proportional to the Yukawa
coupling matrices for up- and down-type quarks, respectively.
Note that we neglect the masses of the light quarks, and therefore we
set the Yukawa couplings of the light quarks to zero.

Since we are interested in the $b \to s\gamma$ transition, we find it
convenient to define
\begin{equation}
\label{def:SUSY-FCNC}
 {\cal G}^{jal} = \frac{G^{jal}}{K_{\mathrm{t}l}}
 \quad \mbox{for $l=b,s$} ; \qquad
 {\cal H}^{\jab} = \frac{H^{\jab}}{K_\mathrm{tb}}
\end{equation}
Unitarity of the mixing (\ref{def:squark-mixing}) implies that
\begin{equation}
\label{eq:squark:unitarity}
\sum_{a=1}^6 \Gamma_{Q\L,\R}^{ai} \Gamma_{Q\L,\R}^{*ak} = \delta_{ik}
\quad , \quad
\sum_{a=1}^6 \Gamma_{Q\L,\R}^{ai} \Gamma_{Q\R,\L}^{*ak} = 0  \; .
\end{equation}
and therefore
\begin{equation}
\label{eq:squark:super-GIM}
\sum_{a=1}^6 {\cal G}^{*\jas} {\cal G}^{\jab} = \lambda_\mathrm{t}^2 |V_{j2}|^2
\quad , \quad
\sum_{a=1}^6 {\cal G}^{*\jas} {\cal H}^{\jab} = 0 \; .
\end{equation}

After having described our conventions, let us now turn to the
evaluation of the QCD correction.  As the squarks and the charginos can
have large mass splittings, the procedure of matching and running will
become more involved but still remains straightforward.  We will give
all ingredients, but the precise procedure will depend on the details of
the spectrum.

%%%%%%%%%%%%%%%%%%%%%%%%%%%%%%%%%%%%%%%%%%%%%%%%%%%%%%%%%%%%%%%%%%%%%%%%

\subsection{Effective operators from heavy squarks}

In the process of evolving down, if we encounter the threshold of an
up-type squark $\tilde u_a$, we will integrate it out.  This generates
effective four-fermion operators made out of the quarks $b$, $s$, and
the active charginos $\chi_j$.  We extend our operator basis by the
following operators (no sum over $j$):
\begin{eqnarray}
 W_\LR^{1,j} & = &
 \mu^{2\eps} \mynorm{g_3^2} \frac{\mb}{\tilde m_j}
 (\bar s_\L b_\R) (\bar\chi_\L^j \chi_\R^j)
\nonumber \\
 W_\LR^{2,j} & = &
 \quarter \mu^{2\eps} \mynorm{g_3^2} \frac{\mb}{\tilde m_j}
 (\bar s_\L \sigma^{\mu\nu} b_\R) (\bar\chi_\L^j \sigma_{\mu\nu} \chi_\R^j)
\nonumber \\
 W_\L^{j} & = &
 \mu^{2\eps} \mynorm{g_3^2}
 (\bar s_\L \gamma^\mu  b_\L) (\bar\chi_\R^j \gamma_\mu \chi_\R^j)
\end{eqnarray}
The matching contributions at $\mu=\tilde m_a$ for $\tilde m_j \gg
\tilde m_a$ are:
\begin{eqnarray}
\label{Matching:msquark:squark}
\Deltaq C_{\Oone} & = & {\cal G}^{*\jas} {\cal H}^{\jab}
\cdot \frac{\tilde m_j}{\mb} \left(\frac{\mW}{\tilde m_a}\right)^2
\cdot (-1)
\nonumber \\
\Deltaq C_{\Otwo} & = & 0
\nonumber \\
\Deltaq C_{\Othree} & = &
 {\cal G}^{*\jas} {\cal H}^{\jab} \cdot
 \frac{\tilde m_j}{\mb} \left(\frac{\mW}{\tilde m_a}\right)^2
  \cdot \frac{(-1)}{\Qb}
\nonumber \\
\Deltaq C_{\Poneone} & = &
 \Deltaq C_{\Ponethree} =
{\cal G}^{*\jas} {\cal G}^{\jab} \cdot
\left(\frac{\mW}{\tilde m_a}\right)^2  \cdot
\left( \frac{5}{18} \right)
\nonumber \\
\Deltaq C_{\Ponetwo} & = &
{\cal G}^{*\jas} {\cal G}^{\jab} \cdot
\left(\frac{\mW}{\tilde m_a}\right)^2  \cdot
\left( -\frac{2}{9} \right)
\nonumber \\
\Deltaq C_{\Ponefour} & = & 0
\nonumber \\
\Deltaq C_{\Ptwo} & = &
{\cal G}^{*\jas} {\cal G}^{\jab} \cdot
\left(\frac{\mW}{\tilde m_a}\right)^2  \cdot
 \frac{1}{2\Qb}
\nonumber \\
\Deltaq C_{\Pfour} & = &
{\cal G}^{*\jas} {\cal G}^{\jab} \cdot
\left(\frac{\mW}{\tilde m_a}\right)^2  \cdot
 \frac{1}{\Qb}
\nonumber \\
\Deltaq C_{W_\LR^{1,j}} & = &
\Deltaq C_{W_\LR^{2,j}} =
{\cal G}^{*\jas} {\cal H}^{\jab} \cdot \frac{16\pi^2}{g_3^2}
\frac{\tilde m_j}{\mb} \left(\frac{\mW}{\tilde m_a}\right)^2
\nonumber \\
\Deltaq C_{W_\L^{j}} & = &
{\cal G}^{*\jas} {\cal G}^{\jab} \cdot \frac{16\pi^2}{g_3^2}
 \left(\frac{\mW}{\tilde m_a}\right)^2
 \cdot (-1)
\end{eqnarray}
The mixing back of these operators into the $O$'s and $P$'s is found to
be:
\begin{equation}
\hat\gamma =
\bordermatrix{
            & O_\LR^{1,2} & \Othree & P_\L^{1,A} & \Ptwo   & \Pfour \cr
W_\LR^{1,j} & 0           & 0       & 0          & 0       & 0 \cr
W_\LR^{2,j} & 0           & -1/\Qb  & 0          & 0       & 0 \cr
W_\L^{j}    & 0           & 0       & 0          & -2/3\Qb & 0 \cr
}
\end{equation}
Since the charginos carry no color charge, the renormalization of these
operators is particularly simple,
\begin{equation}
\hat\gamma =
\bordermatrix{
            & W_\LR^{1,j} & W_\LR^{2,j}       & W_\L^{j} \cr
W_\LR^{1,j} & -2b         & 0                 & 0        \cr
W_\LR^{2,j} & 0           & \frac{16}{3} - 2b & 0        \cr
W_\L^{j}    & 0           & 0                 & -2b      \cr
} \; ,
\end{equation}
and there is no mixing of the $O$ and $P$ operators back into these.

If we cross the threshold of chargino $\chi_j$ at $\mu=\tilde m_j$, the
operators $W^j$ will just be removed; they do not give any matching
contribution to leading order.

%%%%%%%%%%%%%%%%%%%%%%%%%%%%%%%%%%%%%%%%%%%%%%%%%%%%%%%%%%%%%%%%%%%%%%%%

\subsection{Effective operators from heavy charginos}

Let us now consider the case that we encounter the threshold of chargino
$\chi_j$ at $\mu=\tilde m_j$.  If there are still active up-type squarks
$\tilde u_a$, we have to extend our operator basis by the
2-quark--2-squark operators (no sum over $a$):
\begin{eqnarray}
\noalign{\hbox{Dimension $d+1$:}}
\label{def:2quark-2squark}
\tilde Q_\LR^1 & = & \mb \,
 \tilde u_a^{\dagger\beta} \tilde u_a^\alpha \; \bar s_\L^\alpha b_\R^\beta
\nonumber \\
\tilde Q_\LR^2 & = & \mb \,
 \tilde u_a^{\dagger\beta} \tilde u_a^\beta \; \bar s_\L^\alpha b_\R^\alpha
\nonumber \\
\noalign{\hbox{Dimension $d+2$:}}
\Ronel & = & i
 \tilde u_a^{\dagger\beta} \tilde u_a^\alpha
 \left( \bar s_\L \Slash D b_\R \right)^{\alpha\beta}
\nonumber \\
\Rtwol & = & i
 \tilde u_a^{\dagger\beta} (D^\mu\tilde u_a)^\alpha \;
 \bar s_\L^\alpha \gamma_\mu b_\R^\beta
\nonumber \\
\Rthreel & = & i
 (D^\mu \tilde u_a)^{\dagger\beta} \tilde u_a^\alpha \;
 \bar s_\L^\alpha \gamma_\mu b_\R^\beta
\nonumber \\
\Rfourl & = & i
 \tilde u_a^{\dagger\beta} \tilde u_a^\beta \;
 \mathrm{Tr} \, \left(\bar s_\L \Slash D b_\R \right)
\nonumber \\
\Rfivel & = & i
 \tilde u_a^{\dagger\beta} (D^\mu\tilde u_a)^\beta \;
 \bar s_\L^\alpha \gamma_\mu b_\R^\alpha
\nonumber \\
\Rsixl & = & i
 (D^\mu \tilde u_a)^{\dagger\beta} \tilde u_a^\beta \;
 \bar s_\L^\alpha \gamma_\mu b_\R^\alpha
\end{eqnarray}
For $a$ running over each active up-type squarks and using $z_{ja} =
(\tilde m_j/\tilde m_a)^2$, we find the following leading matching
contributions (i.e.\ for large $z_{ja}$) at $\mu=\tilde m_j$:
\begin{eqnarray}
\label{Matching:mchi:chi}
\Deltachi C_{\Oone} & = & {\cal G}^{*\jas} {\cal H}^{\jab} \cdot
 \frac{\tilde m_j}{\mb} \left(\frac{\mW}{\tilde m_j}\right)^2
  \cdot (-1)
\nonumber \\
\Deltachi C_{\Otwo} & = & 0
\nonumber \\
\Deltachi C_{\Othree} & = & {\cal G}^{*\jas} {\cal H}^{\jab} \cdot
 \frac{\tilde m_j}{\mb} \left(\frac{\mW}{\tilde m_j}\right)^2
  \cdot \frac{1}{\Qb}
\nonumber \\
\Deltachi C_{\Poneone} & = &
 \Deltachi C_{\Ponethree} = {\cal G}^{*\jas} {\cal G}^{\jab} \cdot
 \left(\frac{\mW}{\tilde m_j}\right)^2  \cdot
 \left( -\frac{5}{18} \right)
\nonumber \\
\Deltachi C_{\Ponetwo} & = & {\cal G}^{*\jas} {\cal G}^{\jab} \cdot
 \left(\frac{\mW}{\tilde m_j}\right)^2  \cdot
 \left( \frac{11}{9} \right)
\nonumber \\
\Deltachi C_{\Ponefour} & = & 0
\nonumber \\
\Deltachi C_{\Ptwo} & = & {\cal G}^{*\jas} {\cal G}^{\jab} \cdot
 \left(\frac{\mW}{\tilde m_j}\right)^2  \cdot
 \left( -\frac{3}{2} \right) \frac{1}{\Qb}
\nonumber \\
\Deltachi C_{\Pfour} & = & {\cal G}^{*\jas} {\cal G}^{\jab} \cdot
 \left(\frac{\mW}{\tilde m_j}\right)^2  \cdot \frac{1}{\Qb}
\nonumber \\
\Deltachi C_{\tilde Q_\LR^1} & = &
 {\cal G}^{*\jas} {\cal H}^{\jab} \cdot
 \frac{\tilde m_j}{\mb} \left(\frac{\mW}{\tilde m_j}\right)^2 \cdot (-2)
\nonumber \\
\Deltachi C_{\Ronel} & = & \Deltachi C_{\Rtwol} =
 {\cal G}^{*\jas} {\cal G}^{\jab} \cdot
 \left(\frac{\mW}{\tilde m_j}\right)^2 \cdot 2
\nonumber \\
\Deltachi C_{\tilde Q_\LR^2} & = &
 \Deltachi C_{\tilde R_\L^{n,a}} = 0  \qquad n=3,4,5,6
\end{eqnarray}
A straightforward calculation gives for the back-mixing at one-loop
order (but order $\alpha_3^0$ in our chosen normalization)
\begin{equation}
\label{mixing:P:tildeQR}
\gamma^{(0)} =
\bordermatrix{
        & O_\LR   & \Poneone    & \Ponetwo & \Ponethree  & \Ponefour
        & \Ptwo   & \Pfour \cr
\tilde Q_\LR^1 & 0 & 0          & 0        & 0           & 0
        & 0       & 0 \cr
\tilde Q_\LR^2 & 0 & 0          & 0        & 0           & 0
        & 0       & 0 \cr
\Ronel  & 0       & 0           & 0        & 0           & 0
        & 0       & 0 \cr
\Rtwol  & 0       & \frac{1}{6} & -\third  & \frac{1}{6} & 0
        & \frac{1}{6\Qb} & 0 \cr
\Rthreel& 0       &-\frac{1}{6} &  \third  &-\frac{1}{6} & 0
        &-\frac{1}{6\Qb} & 0 \cr
\Rfourl & 0       & 0           & 0        & 0           & 0
        & 0       & 0 \cr
\Rfivel & 0       & 0           & 0        & 0           & 0
        & \myhalf \QtQb & 0 \cr
\Rsixl  & 0       & 0           & 0        & 0           & 0
        &-\myhalf \QtQb & 0
}
\end{equation}
Again one sees that that, similarly to the case of the four-quark
operators, the mixing into the magnetic moment operators vanishes after
applying the equations of motion.  Therefore we have to consider this
mixing at two-loop order.

\begin{figure}[bht]
  \begin{center}
    \epsfxsize=135mm
    \leavevmode
    \epsffile{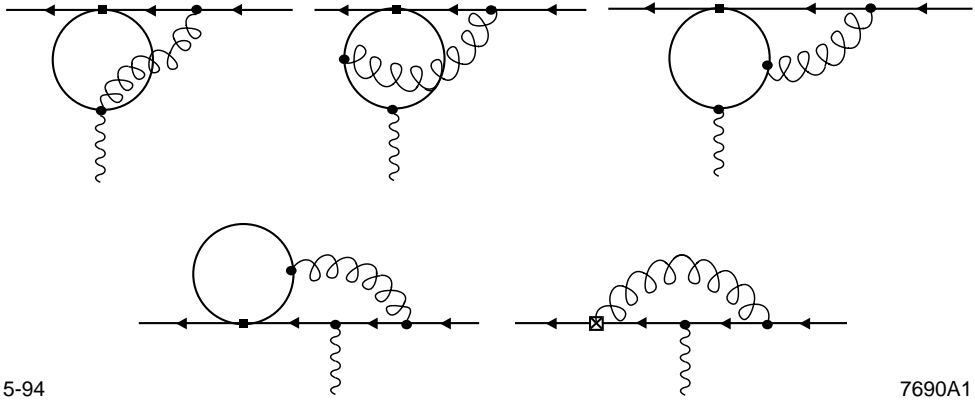}
    \caption{Feynman diagrams contributing to the mixing of the
      two-quark--two-squark operators (\protect\ref{def:2quark-2squark})
      into the operator $\Othree$.
      A full square denotes the insertion of a two-quark--two-squark
      operator, while an open square denotes a one-loop counterterm.
      Diagrams which are related to the ones above by reflection are
      not shown.}
    \label{fig:fig1}
  \end{center}
\end{figure}

The actual two-loop calculation of the mixing of the
two-quark--two-squark operators (\ref{def:2quark-2squark}) into the
magnetic moment operators is performed analogously to the corresponding
calculation with insertions of four-quark operators (see e.g.\
\cite{GSW90}).  In figure~\ref{fig:fig1} we show the relevant diagrams
and one-loop counterterms contributing to the mixing of the
two-quark--two-squark operators into the operator $\Othree$.  As we
prefer to work off-shell, we have to consider only 1-PI diagrams.  The
main advantage is a simplification of the extraction of the divergent
parts of interest by focussing on the coefficients of the tensor
structures that are defined by our basis (\ref{eq:LE-Basis}).

\begin{figure}[thb]
  \begin{center}
    \epsfxsize=135mm
    \leavevmode
    \epsffile{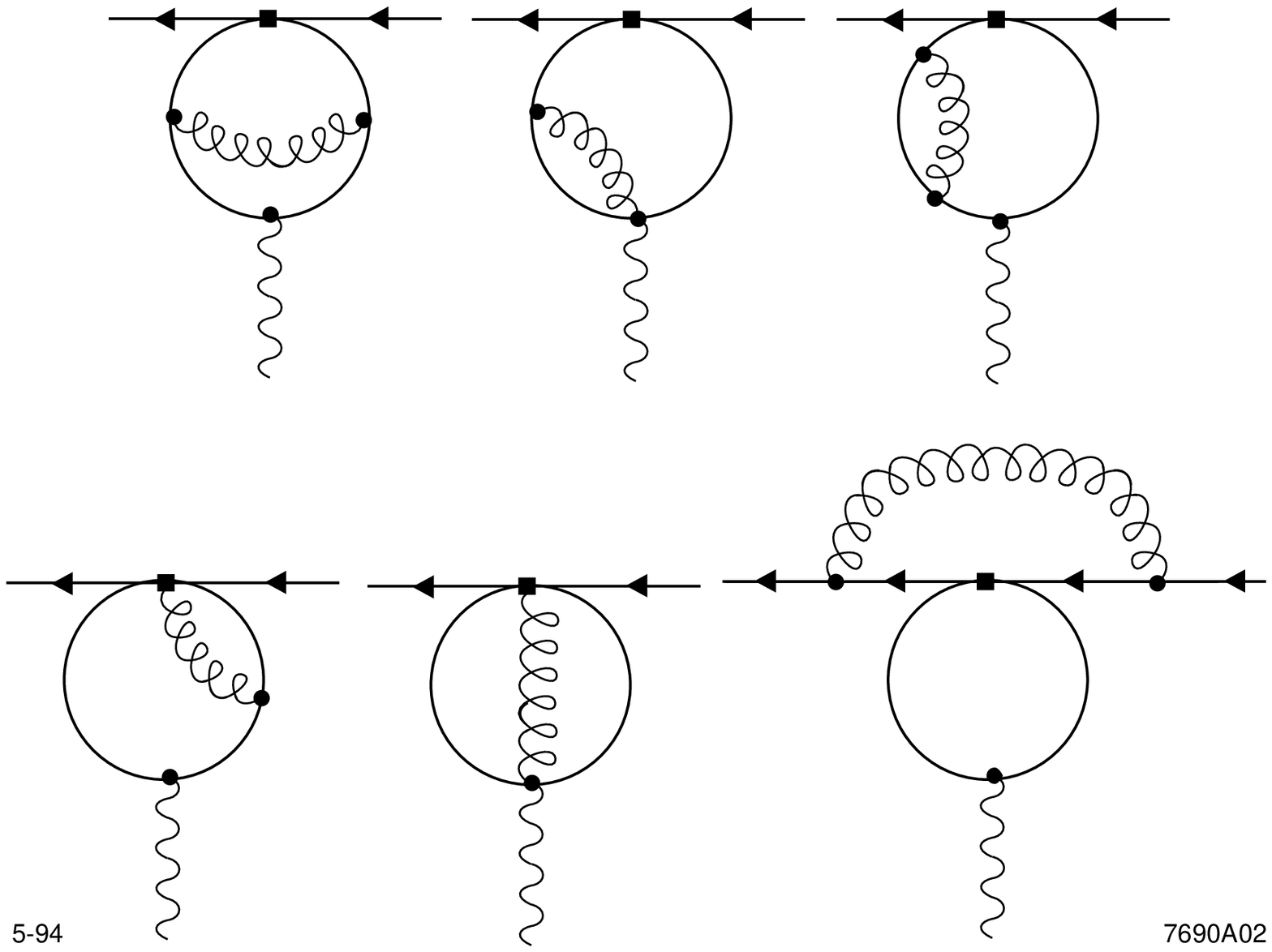}
    \caption{Feynman diagrams whose contribution to the mixing
      vanishes after application of the equations of motion.}
    \label{fig:fig2}
  \end{center}
\end{figure}

Similarly to the corresponding calculations with insertions of
four-fermion operators, using the equations of motion (\ref{eq:EOM})
greatly reduces the computational effort.  Figure~\ref{fig:fig2} shows
typical diagrams which do not contribute because their sum can be shown
to be proportional to $(\gamma_\mu q^2 - q_\mu \slash{q})$, and
therefore need not be calculated.

\begin{figure}[thb]
  \begin{center}
    \epsfxsize=135mm
    \leavevmode
    \epsffile{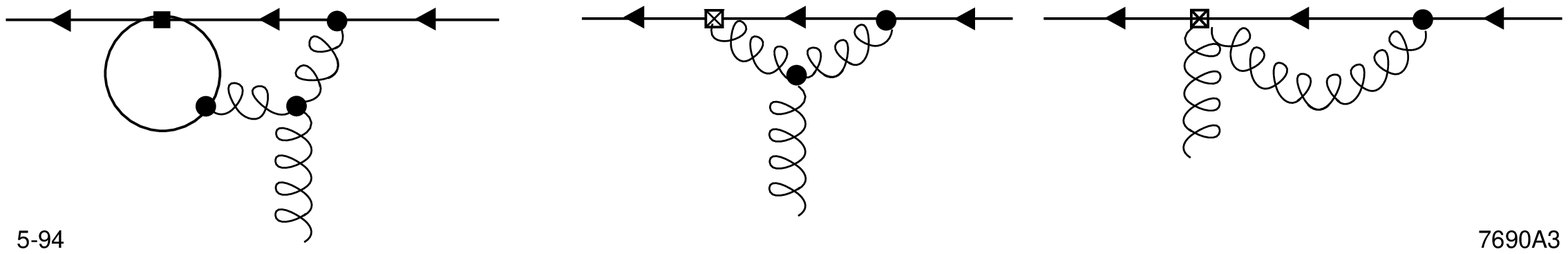}
    \caption{Additional Feynman diagrams which contribute to the mixing
      of the two-quark--two-squark operators into the operator $\Otwo$.}
    \label{fig:fig3}
  \end{center}
\end{figure}

In the case of the mixing into $\Otwo$, due to the non-Abelian
interactions of the gluons, we have to consider the additional diagrams
and counterterms shown in figure~\ref{fig:fig3}.

We obtained the following mixing coefficients ($N=3$):
\begin{equation}
\hat\gamma =
\bordermatrix{
               & \Otwo               & \Othree                \cr
\tilde Q_\LR^1 & \frac{N^2 - 2}{8 N} & \QtQb \frac{N^2-1}{4N} \cr
\tilde Q_\LR^2 & \quarter            & 0                      \cr
\Ronel         & \frac{N^2 - 2}{8 N} & \QtQb \frac{N^2-1}{4N} \cr
\Rtwol         &
       -  \frac{N^2 - 2}{16 N} - \frac{N^2 + 2}{72 N} &
       \left( -\quarter\QtQb + \frac{1}{18} \right) \frac{N^2-1}{2N} \cr
\Rthreel       &
       -  \frac{N^2 - 2}{16 N} + \frac{N^2 + 2}{72 N} &
       \left( -\quarter\QtQb - \frac{1}{18} \right) \frac{N^2-1}{2N} \cr
\Rfourl        & \quarter            & 0                    \cr
\Rfivel        & -\frac{1}{8}        & 0                    \cr
\Rsixl         & -\frac{1}{8}        & 0                    \cr
}
\end{equation}
In addition we need the mixing among the two-quark--two-squark
operators, where the squarks are of the same kind,
\begin{equation}
\hat\gamma =
\bordermatrix{
        &\tilde Q_\LR^1&\tilde Q_\LR^2& \Ronel       & \Rtwol
        & \Rthreel     & \Rfourl      & \Rfivel      & \Rsixl \cr
\tilde Q_\LR^1& \myhalf  & -\frac{3}{2} & 0            & 0
        & 0            & 0            & 0            & 0 \cr
\tilde Q_\LR^2&  0     & -4           & 0            & 0
        & 0            & 0            & 0            & 0 \cr
\Ronel  &  \frac{9}{2} & -\frac{3}{2} & -4           & 0
        & 0            & 0            & 0            & 0 \cr
\Rtwol  & -\frac{1}{2} &  \frac{3}{2}
%  & 0 & -\frac{17}{4}+\frac{1}{12} &  \frac{1}{4}-\frac{1}{12}
%  & 0 &  \frac{3}{4} -\frac{1}{36} & -\frac{3}{4}+\frac{1}{36} \cr
 & 0 &-\frac{25}{6} & \frac{1}{6}  & 0 & \frac{13}{18} &-\frac{13}{18} \cr
\Rthreel& -4           & 0
%  & 0 &  \frac{1}{4} -\frac{1}{12} & -\frac{17}{4}+\frac{1}{12}
%  & 0 & -\frac{3}{4} +\frac{1}{36} &  \frac{3}{4}-\frac{1}{36} \cr
 & 0 & \frac{1}{6}  &-\frac{25}{6} & 0 &-\frac{13}{18} & \frac{13}{18} \cr
\Rfourl & 0            & 0            & 0            & 0
        & 0            & -4           & 0            & 0 \cr
\Rfivel &  \frac{3}{2} & -\frac{1}{2} & 0            & 0
        & 0            & 0            & -2           & -2 \cr
\Rsixl  & -\frac{3}{2} &  \frac{1}{2} & 0            & 0
        & 0            & 0            & -2           & -2  \cr
}
\end{equation}
and for different types of squarks ($a \ne b$):
\begin{equation}
\label{eq:ADM:Ra-Rb}
\hat\gamma =
\bordermatrix{
&\tilde R_\L^{2,b}&\tilde R_\L^{3,b}&\tilde R_\L^{5,b}&\tilde R_\L^{6,b}\cr
\Rtwol
& \frac{1}{12} &-\frac{1}{12} &-\frac{1}{36} & \frac{1}{36} \cr
\Rthreel
&-\frac{1}{12} & \frac{1}{12} & \frac{1}{36} &-\frac{1}{36} \cr
}
\end{equation}
Note that there is a mixing of some of the two-quark--two-squark
operators into four-fermion operators:
\begin{equation}
\label{ADM:Rtilde:S}
\hat\gamma =
\bordermatrix{
        & S_3          & S_4          & S_5          & S_6 \cr
\Rtwol  &-\frac{1}{36} & \frac{1}{12} &-\frac{1}{36} & \frac{1}{12} \cr
\Rthreel& \frac{1}{36} &-\frac{1}{12} & \frac{1}{36} &-\frac{1}{12} \cr
}
\end{equation}
If there are squarks lighter than the top quark, we also have to take
into account the mixing of the four-fermion operators into the operators
$\tilde R$:
\begin{equation}
\label{ADM:S:Rtilde}
\hat\gamma =
\bordermatrix{
    & \Rtwol        & \Rthreel       & \Rfivel        & \Rsixl \cr
S_1 & 0             & 0              & 0              & 0             \cr
S_2 & \frac{1}{3}   & -\frac{1}{3}   & -\frac{1}{9}   & \frac{1}{9}   \cr
S_3 & \frac{2}{3}   & -\frac{2}{3}   & -\frac{2}{9}   & \frac{2}{9}   \cr
S_4 & \frac{n_f}{3} & -\frac{n_f}{3} & -\frac{n_f}{9} & \frac{n_f}{9} \cr
S_5 & 0             & 0              & 0              & 0             \cr
S_6 & \frac{n_f}{3} & -\frac{n_f}{3} & -\frac{n_f}{9} & \frac{n_f}{9} \cr
}
\end{equation}
In principle there is also a QCD-induced mixing into operators with two
quarks and two down-type squarks, which we also would have to include if
we were considering the contributions induced by gluinos and
neutralinos.  In most scenarios, the mass splitting of down-type squarks
is much smaller than for up-type squarks.  For the supersymmetric
contributions to be numerically relevant the lightest squark (which is
usually the lightest stop) must be significantly lighter than the other
squarks.  As has been argued in the introduction, contributions from
these operators are strongly suppressed, and also inclusion of these
operators into the mixing would lead to only a minor effect compared to
other neglected corrections.  Furthermore, all Wilson coefficients that
contribute to mixing via (\ref{eq:ADM:Ra-Rb},\ref{ADM:Rtilde:S}) are
proportional to $\cot^2\beta$ and therefore suppressed in the
large-$\tan\beta$ limit.

Again, if we cross the threshold $\mu=\tilde m_a$ of squark $\tilde
u_a$, the matching contribution vanishes to leading order, and the
operators $\tilde O^a$, $\tilde R^a$ are simply removed.  We will also
add in the corresponding subleading contributions each time a pair
($a,j$) of squarks and charginos has been integrated out, i.e.\ at $\mu
= \min(\tilde m_a, \tilde m_j)$.

%%%%%%%%%%%%%%%%%%%%%%%%%%%%%%%%%%%%%%%%%%%%%%%%%%%%%%%%%%%%%%%%%%%%%%%%

\section{Results and Discussions}

As the full anomalous dimension matrix is quite large and changes its
structure every time we cross a threshold, it would be a big effort to
diagonalize it in every step.  It is much simpler to directly evaluate
the solution (\ref{eq:RGE-solution}) of the RGE numerically.  Before we
proceed, let us comment on some simplifications that result from the use
of the equations of motion, since we are eventually only interested in
the coefficient of the magnetic moment operators at the $b$ scale.

First we note that the operators $Q_\LR$ and $\Rone$, and in the case of
$\mH < \mt$ their ``primed'' counterparts, which appear in intermediate
stages of the calculations, turn out to be superfluous as they do not
give any contribution in the process of matching, nor do they mix into
any other operator.  Second, although the coefficient of $\Ptwo$ does
get matching contributions, and many operators mix into it, it can be
ignored, since it vanishes after applying the equations of motion.
Third, the operators $\Rtwo$ and $\Rthree$ mix only into $\Ptwo$, which
vanishes by equations of motion, and may therefore be omitted from the
beginning.  Extending this reasoning to $\Rtwo'$, $\Rthree'$,
$W_\LR^{1,j}$, $W_\L^j$ and $\Rfourl$ shows that they may also be
disregarded.

Next, one may convince oneself that the apparent zeroth-order mixing of
some operators (see eqs.~(\ref{mixing:P:S1-S6}),
(\ref{mixing:P:tildeQR})) vanishes after application of the equations of
motion, so all mixing occurs at order $(g_3^2/8\pi^2)$, as promised.

% ----------------------------------------------------------------------

Let us first rediscuss the effect of the QCD corrections to the Standard
Model contribution.  For the contribution from the W-$t$-loop there is a
QCD enhancement of the coefficients $C_{\Otwo}^\mathrm{eff}(\mW)$ and
$C_{\Othree}^\mathrm{eff}(\mW)$ of the order of 10--18\% and 15--22\%
for $\mt = 130\ldots 250 \, \GeV$, respectively, which after scaling
down to $\mu=\mb$ and including the contribution from the four-fermion
operators leads to an additional enhancement of the decay rate within
the SM of the order of 12--23\% \cite{CG91}, compared to the case when
both $t$ and W are integrated out at $\mu=\mW$.  This large correction,
which seems to compare quite well with the naive estimate given in the
introduction, is a confirmation that a full next-to-leading order
calculation is quite important.

The magnitude of this effect may be understood by solving the
renormalization group equation for the leading terms.  After application
of the equations of motion, their contribution turns out to be quite
simple:
\begin{eqnarray}
\label{eq:corr:SM}
  \left. C_\Otwo^\mathrm{eff}(\mW) \right|_\mathrm{SM} & = &
 - \frac{5}{24} \left( \frac{\alpha_3(\mt)}{\alpha_3(\mW)} \right)^{14/23}
 + \frac{1}{3} + \mathrm{(subleading)}
\nonumber \\
  \left. C_\Othree^\mathrm{eff}(\mW) \right|_\mathrm{SM} & = &
   \frac{5}{3}  \left( \frac{\alpha_3(\mt)}{\alpha_3(\mW)} \right)^{14/23}
 - \frac{8}{3} + \mathrm{(subleading)}
\end{eqnarray}
The reason that (\ref{eq:corr:SM}) leads to positive corrections is
essentially due to the fact that the effective matching contributions at
$\mu=\mt$ (\ref{Matching:mt:SM}) and at $\mu=\mW$ (\ref{Matching:mw:SM})
have opposite sign (which is a remnant of the GIM mechanism), and
therefore lead to coefficients of opposite sign but comparable magnitude
of the first two terms on the right-hand sides of (\ref{eq:corr:SM}).
It has long been known \cite{GW80} that the QCD corrections tend to
soften the GIM-cancellations between different up-type quarks if they
are nearly degenerate; but even for a heavy top quark (i.e.\ $\mt \gg
\mW$) there remains a finite enhancement, as can be explicitly seen from
these expressions.  Note that (\ref{eq:corr:SM}) gives only the leading
terms; the subleading terms being suppressed by only a factor of
$(\mW/\mt)^2$.

% ----------------------------------------------------------------------

Next let us turn to the contribution from the loop with a charged Higgs.
For this case we have solved the renormalization group equation
numerically, using as input parameters:
\begin{eqnarray*}
&&
\mb = 4.5 \, \GeV , \quad
\mt = 175 \, \GeV ,
\\ &&
\mW = 80.22 \, \GeV , \quad
\alpha_3(m_\mathrm{Z}) = 0.123
\end{eqnarray*}

\begin{figure}[tb]
  \begin{center}
    \begin{picture}(135,90)
      \put(0,0){\makebox(135,90){
        \epsfxsize=135mm
        \leavevmode
        \epsffile{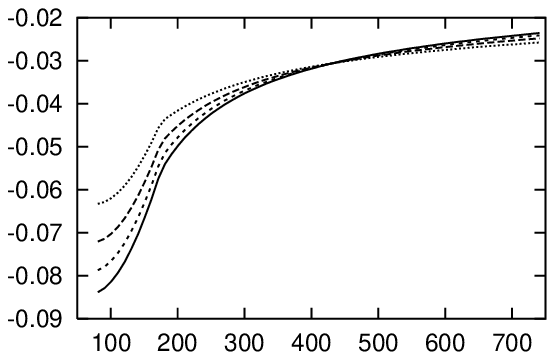}
      }}
      \put(65,0){$\mH$ [GeV]}
      \put(0,47){$\delta_\cHiggs$}
    \end{picture}
  \end{center}
  \caption{Correction (\protect\ref{def:delta_H}) to the coefficients of
    $\Othree(\mb)$ from a loop with $t$ quark and charged Higgs with
    leading QCD corrections from large mass splitting to the case when
    both particles are integrated out at $\mu=\mW$.
    The mass of the $t$ quark is assumed to be 175~GeV.
    The dotted, long-dashed, dashed and solid line correspond to
    $\tan\beta = 1.5, \, 2, \, 3,$ and 10, respectively.}
  \label{fig:H-contrib}
\end{figure}

The resulting correction
\begin{equation}
\label{def:delta_H}
  \delta_\cHiggs = \left.
  \frac{C_\Othree^\mathrm{eff}(\mb)}{C_\Othree^\mathrm{eff,naive}(\mb)}
  \right|_{\cHiggs}
  - 1
\end{equation}
to the ``naive'' result, obtained by integrating out $t$ and $\cHiggs$
simultaneously at the W scale, is shown in figure~\ref{fig:H-contrib}
for $\mW < \mH < 750 \, \GeV$ and $\tan\beta = 1.5, \, 2, \, 3,$ and 10.
At sufficiently large $\tan\beta$ (i.e.\ $\tan\beta > 3$), the
correction turns out to be essentially independent of $\tan\beta$.  This
is quite understandable since the $\tan\beta$-dependent pieces are
actually proportional to $\cot^2\beta$.

For a light charged Higgs, i.e.~$\mH < \mt$, there appears to be a
further reduction of this contribution compared to the naive result.
Indeed, in the limit of large $\tan\beta$, and assuming there is no
light squark or gluino with mass below $\mt$, one finds the following
simple analytical result for the charged Higgs contribution, valid for
$\mb < \mu < \mt$:
\begin{eqnarray}
\label{eq:corr:light-H}
  \left. C_\Otwo^\mathrm{eff}(\mu) \right|_\cHiggs & = &
   \frac{1}{4} \left( \frac{\alpha_3(\mt)}{\alpha_3(\mu)} \right)^{14/23}
 + \mathrm{(subl.)} + O(\cot^2\beta)
 \\
  \left. C_\Othree^\mathrm{eff}(\mu) \right|_\cHiggs & = &
   \frac{3}{4} \left( \frac{\alpha_3(\mt)}{\alpha_3(\mu)} \right)^{16/23}
 - 2           \left( \frac{\alpha_3(\mt)}{\alpha_3(\mu)} \right)^{14/23}
 + \mathrm{(subl.)} + O(\cot^2\beta)
\nonumber
\end{eqnarray}
Hence no enhancement occurs as in the case of the SM contribution; on
the contrary, the leading coefficients get suppressed as they are run
down from the $\mt$, compared to the subleading terms which (according
to our discussion in section \ref{sec:subleading}) get only suppressed
by the evolution from $\mu=\mH$ down to $\mu=\mb$.  For sufficiently
small $\mH$, the additional QCD corrections are then essentially due to
the running from $\mu=\mt$ to $\mu=\mH$.  Note that our ``corrections''
are counted relative to the case when both particles in the loop are
integrated out at the common scale $\mu=\mW$, which is obtained from
(\ref{eq:corr:light-H}) by substituting $\alpha_3(\mt)
\to \alpha_3(\mW)$.  Thus, for $\mH < \mt$, integrating out $t$ and
$\cHiggs$ at the $t$ scale appears to give a more accurate result than
at $\mu=\mH$ or $\mu=\mW$.

On the other hand, for $\mH > \mt$ we found only a minor suppression of
a few percent, which is essentially the result of a partial cancellation
of the enhancement coming from the scaling between $\mH$ and $\mt$ (due
to one negative eigenvalue of the submatrix (\ref{ADM:S7-S10}) for the
mixing of four-fermion operators), and of a reduction from the scaling
between $\mt$ to $\mW$.  Unfortunately, we were unable to obtain a
simple analytical solution for this case.

% ----------------------------------------------------------------------

In the case of the chargino contribution, things are more complicated,
since one has to consider in general the dependence of the amplitude as
a function of several parameters, namely the mass spectrum and the
mixing angles for the charginos and the up-type squarks.  However, it
turns out that the essential features may already be studied for the
case of sufficiently large $\tan\beta$, which is in the center of recent
interest \cite{Bor93,Osh93,Dia93,GN93,HKT94,BV94}.  In this case, the
parameter $\lambda_\mathrm{b}$ (\ref{eq:Yukawa}) may become of the same
order of magnitude as the parameter $\lambda_\mathrm{t}$.  Assuming
furthermore that the mixing in the squark sector is essentially the same
as in the quark sector, which is quite natural in supergravity models
where the soft SUSY-breaking is characterized by a common scalar mass at
some unification scale, the quantities ${\cal G}$ and ${\cal H}$, as
defined in (\ref{def:SUSY-FCNC}), are then necessarily of the same order
of magnitude, the terms proportional to the ratio $\tilde m_j/\mb$ will
dominate the amplitude, and the corrections become
$\tan\beta$-independent.

In this particular limit, one can find an analytical result for the
leading terms.  For the case of the chargino being much lighter than the
squark, the coefficients read:
\begin{eqnarray}
\label{eq:corr:light-chargino}
  \left. C_\Otwo^\mathrm{eff}(\mu) \right|_{\chi} & = &
   \tilde C \cdot
   \frac{1}{2} e^{14 t/3}
 + \mathrm{(subl.)} + O(\cot^2\beta)
\nonumber \\
  \left. C_\Othree^\mathrm{eff}(\mu) \right|_{\chi} & = &
   \tilde C \cdot \left[
   -4 e^{14 t/3}
 + e^{16 t/3} \left(\frac{15}{2} + \frac{4\pi}{2b\Qb} \left(
   \frac{1}{\alpha_3(\mu)} - \frac{1}{\alpha_3(\tilde m_a)}
   \right) \right)  \right] +
\nonumber \\ &&
 + \mathrm{(subl.)} + O(\cot^2\beta)
 \\
\noalign{\hbox{where}}
 \tilde C & = & {\cal G}^{*\jas} {\cal H}^{\jab}
  \cdot \frac{\tilde m_j}{\mb} \left(\frac{\mW}{\tilde m_a}\right)^2
 \; , \qquad
 t =
  \frac{1}{2b} \ln \left( \frac{\alpha_3(\mu)}{\alpha_3(\tilde m_a)} \right)
\nonumber
\end{eqnarray}
while for the other case of a squark much lighter than a chargino, we
get:
\begin{eqnarray}
\label{eq:corr:light-squark}
  \left. C_\Otwo^\mathrm{eff}(\mu) \right|_{\tilde q} & = &
  \tilde C \cdot \left[
   \frac{99}{260} e^{14 t/3}
 + \frac{1}{10}   e^{t/2}
 + \frac{1}{52}   e^{-4 t}
  \right]
 + \mathrm{(subl.)} + O(\cot^2\beta)
\nonumber \\
  \left. C_\Othree^\mathrm{eff}(\mu) \right|_{\tilde q} & = &
  \tilde C \cdot \left[
 - \frac{198}{65}  e^{14 t/3}
 + \frac{495}{406} e^{16 t/3}
 - \frac{96}{145}  e^{t/2}
 - \frac{1}{91}    e^{-4 t}
  \right] +
\nonumber \\ &&
 + \mathrm{(subl.)} + O(\cot^2\beta)
 \\
\noalign{\hbox{where now}}
 \tilde C & = & {\cal G}^{*\jas} {\cal H}^{\jab}
  \cdot \frac{\mW^2}{\tilde m_j \mb}
 \; , \qquad
 t =
  \frac{1}{2b} \ln \left( \frac{\alpha_3(\mu)}{\alpha_3(\tilde m_j)} \right)
\nonumber
\end{eqnarray}
At first sight the terms proportional to $1/\alpha_3$ in
(\ref{eq:corr:light-chargino}) might be embarrassing, but a closer look
shows that their difference is (to leading order) just some number times
$\ln(\mu/\tilde m_a)$ and therefore finite in the limit $\alpha_3 \to 0$.

Unfortunately, the interpretation of these expressions is aggravated in
both limiting cases since the number of free parameters in the general
model is quite large, and due to eqs.~(\ref{eq:squark:super-GIM}) one
has a supersymmetric version of the GIM mechanism, which leads to a
partial cancellation of the leading terms under consideration.  This
renders it difficult to estimate the actual corrections due to the mass
splitting between charginos and squarks by using
(\ref{eq:corr:light-chargino}) or (\ref{eq:corr:light-squark}).

Some features of these expressions may still be studied under the
following assumptions: i) the squarks of the first two generations are
degenerate with mass $\tilde m_u$, ii) the mixing in the squark sector
is the same as in the quark sector (i.e.\ the gluino-quark-squark
couplings are flavor-diagonal even in the mass eigenstate basis), and
iii) the mass matrix for the stop is given (in the ($t_\L, t_\R$) basis)
by
\begin{eqnarray}
\label{eq:stop-mass-matrix}
  M_{\tilde t}^2 & = &
  \pmatrix{
     \tilde m^2_{t_\L}  & \tilde m^2_{t_\LR} \cr
     \tilde m^2_{t_\LR} & \tilde m^2_{t_\R}  \cr
          }
\end{eqnarray}
This mass matrix is diagonalized by a unitary matrix $T$,
\begin{equation}
  T M_{\tilde t}^2 T^{-1} =
  \pmatrix{ \tilde m_{t_1}^2 & 0 \cr  0 & \tilde m_{t_2}^2 \cr }
\end{equation}
In this scenario, the quantities (\ref{def:SUSY-FCNC}) take a
particularly simple form:
\begin{eqnarray}
  {\cal G}^{jal} & \simeq &
    V_{j1} T_{a1} - \lambda_\mathrm{t} V_{j2} T_{a2}
    \qquad \mathrm{for} \; l = b,s; \; a = \tilde t_{1,2}
  \nonumber \\
  {\cal H}^\jab  & \simeq &  \lambda_\mathrm{b} U_{j2} T_{a1}
\end{eqnarray}
while the sum over the squarks of the first two generations is
determined by (\ref{eq:squark:super-GIM}).

Let us for the moment neglect the mixing between $\tilde t_\L$ and
$\tilde t_\R$, i.e.\ consider the case $\tilde m^2_{t_\LR} = 0$,
$T={\bf 1}$.
Evaluating the first line of (\ref{eq:corr:light-chargino}) to lowest
order, we find for the contribution of a light chargino and after
summing over the different squarks:
\begin{eqnarray}
\label{eq:leading-O2eff}
  C_\Otwo^\mathrm{eff}(\tilde m_j) & = &
  \myhalf \lambda_\mathrm{b} U_{j2} V_{j1}
  \frac{\tilde m_j}{\mb} \frac{\mW^2}{\tilde m^2_u}
  \left( \frac{\tilde m^2_j}{\tilde m^2_u} \right)^{
     \frac{14}{3} \hat\alpha_3 }
  \left[ \left( \frac{\tilde m^2_u}{\tilde m^2_{t_\L}} \right)^{
     1 + \frac{14}{3} \hat\alpha_3 }
   - 1 \right] +
\nonumber \\ &&
   + O(\hat\alpha_3^2) + \mathrm{(subl.)} + O(\cot^2\beta)  \\
\noalign{\hbox{with the abbreviation}}
  \hat\alpha_3 & = & \frac{\alpha_3(\tilde m_j)}{4\pi}
\nonumber
\end{eqnarray}
Similar, although rather lengthy expressions are obtained if the mixing
between $\tilde t_\L$ and $\tilde t_\R$ is taken into account, and
analogous results are found for the other coefficients in
(\ref{eq:corr:light-chargino}) and (\ref{eq:corr:light-squark}).  As has
already been pointed out in \cite{GN93}, the sign of the product $U_{j2}
V_{j1}$ depends on the sign of $\muh$, so that this leading contribution
for large $\tan\beta$ can have either sign.

A closer look at (\ref{eq:leading-O2eff}) shows two counteracting
effects: a reduction of the leading coefficient due to QCD running from
$\tilde m_u$ down to $\tilde m_j$, while the term in square brackets
shows an enhancement due to a ``QCD-softening'' of the GIM cancellation,
independent on whether $\tilde m_{t_\L}$ is larger or smaller than
$\tilde m_u$.  The actual size of the corrections depends of course on
the mass splitting between the squarks as well as on the splitting
between the mass of the chargino and the squarks; since squarks can be
an order or magnitude heavier than the lightest chargino, we estimate
this coefficient to be of the order of
\[
  \frac{14}{3} \frac{\alpha_3(\tilde m_j)}{4\pi}  \times
  \left( \ln \frac{\tilde m_u}{\tilde m_j} , \;
         \ln \frac{\tilde m_u}{\tilde m_t} \right)
  \simless 15 \%
\]
Similar results are found when analyzing the other expressions, so we
will in general expect corrections up to $O(15\%)$ with either sign.  An
exceptional situation occurs when, due to these super-GIM cancellations,
the lowest-order contribution to $\Othree$ is accidentally lower than to
$\Otwo$ by orders of magnitude, since the above reasoning did not take
into account the mixing of $\Otwo$ into $\Othree$ for scales below the
heavy thresholds.  In this case a sensible answer is obtained only when
using the full expressions.

%%%%%%%%%%%%%%%%%%%%%%%%%%%%%%%%%%%%%%%%%%%%%%%%%%%%%%%%%%%%%%%%%%%%%%%%

\section{Conclusions}

We have extended the calculation of the leading QCD corrections for the
inclusive $\btosg$ decay to the MSSM in the framework of effective field
theories.  It was shown that it is important to properly treat the
high-energy scale at which the particles in the loop are integrated out,
as well as how to calculate the QCD corrections between if the masses of
the particles in the loop are vastly different.  To this end, we have
calculated the leading order anomalous dimension matrices for the
operators for the various scenarios that are relevant to this process in
the MSSM.

We found that, while the SM contribution to the Wilson coefficients at
the weak scale gets enhanced in the limit of a heavy top quark by about
15--20\%, the contribution from a loop with a charged Higgs gets
actually slightly reduced by a few percent.  For the contribution from
the chargino loops the result depends strongly on the mass spectrum of
the squarks and the charginos as well as on the mixing angles.
Typically, one expects corrections up to the order of 15\% with either
sign, which is however less than the enhancement of the SM contribution.

Given a range of values for the inclusive decay, if one applies the
above results to a parameter space analysis for a particular SUSY model,
one will essentially find a relaxation of the bounds on the mass of the
charged Higgs, especially in the region of large $\tan\beta$.  The
impact of the modification of the QCD corrections for the chargino loop
contribution is not seen so easily, but we expect a smooth deformation
of contours in analyses like \cite{Bor93,BV94}, with the strongest
effect in those regions where the lowest order contribution to the
coefficient $C_\Othree$ is small although the chargino is relatively
light.

Finally we would like to point out that for the inclusive decay rate,
even after taking into account the real gluon emission and virtual
corrections below the $b$ scale \cite{AG91}, the leading order
prediction remains uncertain by about 25\% due to the residual scale
dependence alone \cite{AG93,BMMP93} (for the amplitude it is of the
order 10--15\%).  Once a full next-to-leading order calculation is
available for the SM, it may be combined with the above results to
obtain predictions in the MSSM with comparable precision.

%%%%%%%%%%%%%%%%%%%%%%%%%%%%%%%%%%%%%%%%%%%%%%%%%%%%%%%%%%%%%%%%%%%%%%%%

\section*{Acknowledgements}

We would like to thank S. Brodsky, J. Hewett, and T. Rizzo for
discussions and useful comments on the manuscript.

The present work was stimulated by discussions with B. Grinstein during
a stay at the former Superconducting Super Collider Laboratory.  We
benefitted also from conversations with F. Borzumati and J. Wells.

Finally, we would like to thank P. Cho for sending us the erratum to
\cite{CG91} prior to publication.

%%%%%%%%%%%%%%%%%%%%%%%%%%%%%%%%%%%%%%%%%%%%%%%%%%%%%%%%%%%%%%%%%%%%%%%%

\appendix

\section{Wilson coefficients at one loop}

We quote here the results for the Wilson coefficients at one-loop order
when both particles in the loop are integrated out at a common scale.
These results will be used for the determination of subleading terms.
They also provide an important cross-check for the leading terms
obtained by the calculation in the effective theory, as well as for some
of the entries in the anomalous dimension matrix.

We find it convenient to use the following functions that appear in the
evaluation of the coefficients of the basis operators:
\begin{eqnarray}
\label{def:F_i}
F_1(x) & = & \frac{ x^2 - 5x - 2}{12(x-1)^3} + \frac{x  \ln x}{2(x-1)^4}
\nonumber \\
F_2(x) & = & \frac{2x^2 + 5x - 1}{12(x-1)^3} - \frac{x^2\ln x}{2(x-1)^4}
\nonumber \\
F_3(x) & = & \frac{x - 3}{2(x-1)^2} + \frac{\ln x}{(x-1)^3}
\nonumber \\
F_4(x) & = & \frac{x + 1}{2(x-1)^2} - \frac{x \ln x}{(x-1)^3}
\end{eqnarray}
These functions are identical with those given in the appendix of
ref.~\cite{BBMR91}.  Some of their properties are:
\begin{eqnarray}
&&
F_1 \left(\frac{1}{x}\right) = x F_2(x) , \quad
F_2 \left(\frac{1}{x}\right) = x F_1(x) , \quad
F_4 \left(\frac{1}{x}\right) = x F_4(x)
\nonumber \\ &&
F_1(x) + F_2(x) = \myhalf F_4(x) = \quarter - \myhalf x F_3(x)
\nonumber \\ &&
x F_1(x) + F_2(x) = \frac{1}{12}
\nonumber \\ &&
F_3 \left(\frac{1}{x}\right) =
 -x \left( F_3(x)+ 2F_4(x) \right) + \frac{x \ln x}{x-1}
\end{eqnarray}

% ----------------------------------------------------------------------

\subsection{Standard Model loop contributions}

Integrating out the W, the charged \wbG/s and an up-type quark
simultaneously, we obtain the one-loop expression of the Wilson
coefficients of the effective Hamiltonian (\ref{eq:H_eff}):
\begin{eqnarray}
\label{eq:1-loop:W}
C_{\Oone} & = & - x F_4(x)
\nonumber \\
C_{\Otwo} & = & \frac{\Qb}{\Qt} C_{\Othree} =
 - \frac{x}{2} \left( F_3(x) + F_4(x) \right)
\nonumber \\
C_{\Poneone} & = & C_{\Ponethree} =
 \frac{1}{3} (x+2) \left( 2 F_2(x) + F_3(x) + 2 F_4(x) \right)
\nonumber \\
C_{\Ponetwo} & = &
 \frac{2}{3} (x+2) \left( F_2(x) - F_3(x) - 2 F_4(x) \right)
\nonumber \\
C_{\Ponefour} & = &
 (x-2) F_4(x)
 \\
C_{\Ptwo} & = &
\frac{1}{\Qb} (x+2) \left( \myhalf F_3(x) + F_4(x) - \frac{\ln(x)}{6(x-1)}
 \right)
\nonumber \\
C_{\Pfour} & = &
\frac{1}{\Qb} \left( 3 - 2x F_3(x) - 5x F_4(x) \right)
\nonumber
\end{eqnarray}
Here $x = (m_q/\mW)^2$.  Note that for large $x$ all coefficient
functions are bounded, except for $C_\Ptwo$, which grows logarithmically
with $x$.  For small $x$, $C_\Poneone$, $C_\Ponetwo$, $C_\Ponethree$ and
$C_\Ptwo$ diverge logarithmically.

% ----------------------------------------------------------------------

\subsection{Charged Higgs loop contributions}

Integrating out the charged Higgs and an up-type quark simultaneously,
the corresponding expressions are ($y = (m_q/\mH)^2$):
\begin{eqnarray}
\label{eq:1-loop:H}
C_{\Oone} & = & y F_4(y)
\nonumber \\
C_{\Otwo} & = & \frac{\Qb}{\Qt} C_{\Othree} =
 \frac{y}{2} \left( F_3(y) + F_4(y) \right)
\nonumber \\
C_{\Poneone} & = & C_{\Ponethree} =
 \frac{1}{3} y \left( 2 F_2(y) + F_3(y) + 2 F_4(y) \right) \cot^2\beta
\nonumber \\
C_{\Ponetwo} & = &
 \frac{2}{3} y \left( F_2(y) - F_3(y) - 2 F_4(y) \right) \cot^2\beta
\nonumber \\
C_{\Ponefour} & = & - \Qb C_{\Pfour} =
 y F_4(y) \cot^2\beta
 \\
C_{\Ptwo} & = &
\frac{1}{\Qb} y \left( \myhalf F_3(y) + F_4(y) - \frac{\ln(y)}{6(y-1)}
 \right) \cot^2\beta
\nonumber
\end{eqnarray}

% ----------------------------------------------------------------------

\subsection{Chargino loop contributions}

Finally we give the expressions for integrating out a chargino and an
up-type squark.  Setting $z = (\tilde m_j/\tilde m_a)^2$, where $m_j$
and $m_a$ represent the mass of the chargino $\chi_j$ and of the up-type
squark $\tilde u_a$ respectively, and using the couplings defined in
eq.~(\ref{def:SUSY-FCNC}), one finds
\begin{eqnarray}
C_{\Oone} & = & {\cal G}^{*\jas} {\cal H}^{\jab} \cdot
\frac{\tilde m_j}{\mb} \left(\frac{\mW}{\tilde m_a}\right)^2
\cdot (-2) F_4(z)
\nonumber \\
C_{\Otwo} & = & 0
\nonumber \\
C_{\Othree} & = & {\cal G}^{*\jas} {\cal H}^{\jab} \cdot
\frac{\tilde m_j}{\mb} \left(\frac{\mW}{\tilde m_a}\right)^2
  \cdot \frac{1}{\Qb} \left( F_3(z) + F_4(z) \right)
\nonumber \\
C_{\Poneone} & = & C_{\Ponethree} =
 {\cal G}^{*\jas} {\cal G}^{\jab} \cdot
 \left(\frac{\mW}{\tilde m_a}\right)^2  \cdot \frac{1}{3}
 \left[ 4 F_2(z) + \frac{1}{z} F_3\left(\frac{1}{z}\right) \right]
\nonumber \\
C_{\Ponetwo} & = &
 {\cal G}^{*\jas} {\cal G}^{\jab} \cdot
 \left(\frac{\mW}{\tilde m_a}\right)^2  \cdot \frac{2}{3}
 \left[ 2 F_2(z) - \frac{1}{z} F_3\left(\frac{1}{z}\right) \right]
\nonumber \\
C_{\Ponefour} & = & 0
 \\
C_{\Ptwo} & = &
 {\cal G}^{*\jas} {\cal G}^{\jab} \cdot
 \left(\frac{\mW}{\tilde m_a}\right)^2
 \cdot \left( - \frac{2}{\Qb} \right)
 \left( \myhalf F_3(z) + F_4(z) - \frac{\ln(z)}{6(z-1)} \right)
\nonumber \\
C_{\Pfour} & = &
 {\cal G}^{*\jas} {\cal G}^{\jab} \cdot
 \left(\frac{\mW}{\tilde m_a}\right)^2
\cdot \left( \frac{2}{\Qb} \right)
 F_4(z)
\nonumber
\end{eqnarray}
After application of the equations of motion, these expressions are
consistent with the corresponding expressions in \cite{BBMR91}.

%%%%%%%%%%%%%%%%%%%%%%%%%%%%%%%%%%%%%%%%%%%%%%%%%%%%%%%%%%%%%%%%%%%%%%%%

%\newpage

%%%%%%%%%%%%%%%%%%%%%%%%%%%%%%%%%%%%%%%%%%%%%%%%%%%%%%%%%%%%%%%%%%%%%%%%

\end{document}